# Forecasting the Leading Indicator of a Recession: The 10-Year minus 3-Month Treasury Yield Spread

Author: Sudiksha Joshi,

Pennsylvania State University, University Park, PA[1]

**Abstract**

In this research paper, I have applied various econometric time series and two machine learning models to forecast the daily data on the yield spread − the difference between the 10-year Treasury yields and the 3-month Treasury bills. First, I decomposed the yield curve into its principal components, then simulated various paths of the yield spread using the Vasicek model. After constructing univariate ARIMA models, and multivariate models such as ARIMAX, VAR and Long Short Term Memory (LSTM), I calibrated the root mean squared error (RMSE) to measure how far the models' results deviate from the current values. Through impulse response functions, I measured the impact of various shocks on the difference yield spread. The results indicate that the parsimonious univariate ARIMA model ($RMSE = 0.05185$) outperforms the richly parameterized VAR method ($RMSE = 0.4648$), and the complex LSTM with multivariate data performs equally well as the simple ARIMA model.

[1] For discussion, feedback or comments, please email me at sxj5465@psu.edu

# 1. Introduction

As a leading indicator of predicting a recession, economists typically incorporate the yield spread in probit models to forecast the probability of a recession one year from now. The yield curve depicts the interest rates of treasury securities of various maturities that have equal credit quality and same risk characteristics. The Liquidity Premium Theory of the Term Structure of Interest Rate states that since bondholders face inflation and interest rates risk, the term premium for holding a $n-$year bond compensates the long-term bondholders. A strikingly precise predictor of the ex-ante economic activity such as GDP and industrial production, the yield (or term) spread is the difference between long and short-term interest rates. Hence, I chose the difference between the 10-year and 3-month constant maturity as a measure of yield spread. This is because a negative yield spread, or an inverted yield curve has preceded economic slowdown in the past six decades. Reflecting expectations of future economic conditions, long term rates move up alongside the short-term rates during the initial expansionary periods, and are likely to stop once markets become pessimistic about the economic outlook. A credit crunch may occur once the yield curve flattens, as it is relatively less profitable for banks to borrow for short-term and lend for long-term, dampening the loan supply. Whilst monetary policy directly influences the short-end of the yield curve, they indirectly affect the long-term treasury rate via expectations. A hike in the policy rate at times raises the long-term rates, albeit by a smaller amount than the rise in the short-end of the yield curve as shown in figure 1. Occasionally, they move in the opposite direction or decline without an explicit simultaneous trajectory in the short-term rates as shown in the graph below. In either cases, if the movement persists, then the yield curve inverts.

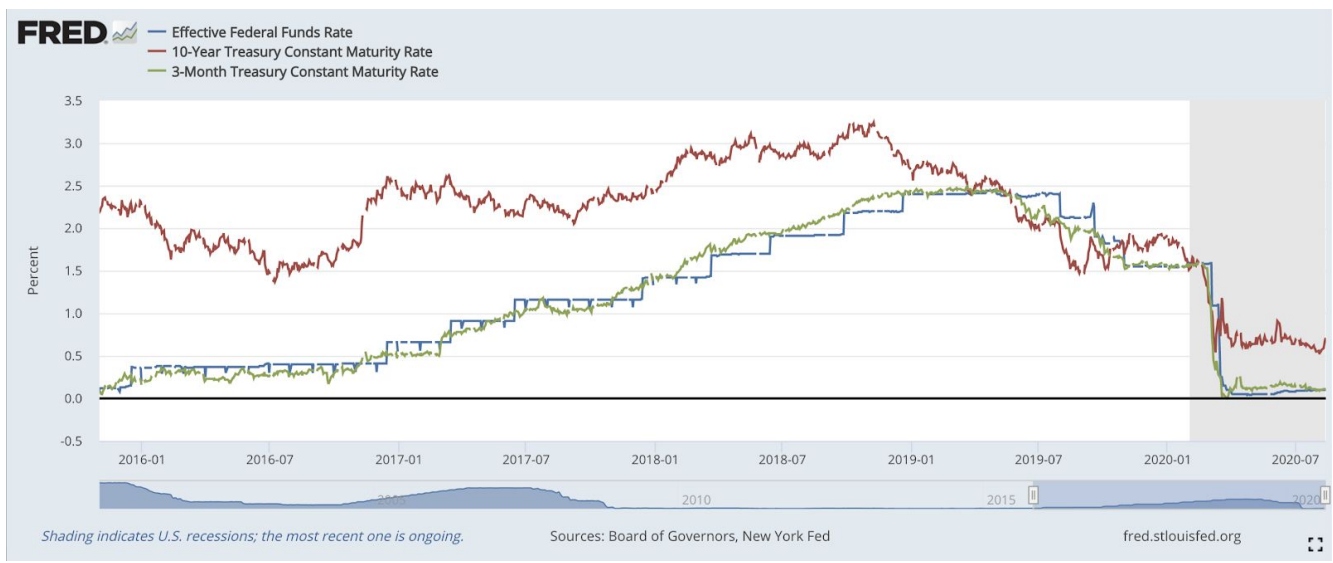

Figure 1: Changes in monetary policy (through federal funds rate) affect the short term greater than the long term Treasury yields

The long term rate constitutes the term or risk premium and expectations of future short-term rates. According to several estimates formulated, including those by Bauer and Rudebusch (2016), the risk premium on long-term interest rates have been close to zero and mostly negative[2] since November 2018 (Kim and Wright 2005). The implications on the economic outlook are typically different if long term rates, and subsequently the term spread, are driven down by lower risk premium, instead of the pessimistic expectations about ex-ante interest rates. However, robust statistical models suggest that both of the components of the term spread contain equal amounts of signal in forecasting the recessions. Moreover, the views on business cycle and monetary policy affect the investors' expectations that shape the expectations of future short-term rates component of the long-term rate. For instance, if investors think that a downturn is imminent, they would expect the FOMC to lower the effective federal funds rate in the future and enhance monetary accommodation. This expectation truncates the long-term rates, inverting the yield curve. Alternatively, markets might think that the current aggressive monetary policy tightening would hike the federal funds rate relative to the policy rates in the future, increasing the odds of an ex-ante slump in the real activity. If the market's projections of a downturn are correct, then a decline in the yield spread would follow a higher probability of an ex-ante recession.

The structure of the research paper is as follows. Section 2 reviews the literature on yield curve and yield spread forecasting. While the literature on yield curve modeling is prolific, not many economists have forecasted yield spread alone, albeit it is derived from the components of the yield curve itself. Section 3 elaborates the models and results presented. First, I have explained how we can decompose the yield curve into its three major various components, including the yield spread. Subsequently, I simulated various trajectories of the yield spread using the Vasicek model, by changing its various parameters. Then, I constructed the forecasts using various methodologies. Finally, I compared the forecast errors from different models and concluded that the simple models such as ARIMA perform better than the reduced-form VAR, but the forecasts from the multivariate LSTM are the most precise. Section 4 concludes and discusses possible ways to enhance the models by employing new methodologies.

## 2. Literature Review

Sharpe and Engstrom (2018) scrutinized an alternative measure of the yield curve slope, known as the "near-term forward spread − difference between the six quarter ahead forward rate on the US Treasury securities and the current three-month Treasury bill." They show that a "near-term forward spread" statistically dominates "long-term spreads", where the former calibrates market's expectations for the near-term path of the federal funds rate. A negative near-term spread connotes that investors' expect the Fed to enforce easing monetary policies in the next 12-18 months as a recession is in the offing. The bonds with maturity greater than eighteen months in maturity have no power in forecasting either the GDP growth rate or a recession. Gauging the slope of the Treasury term structure, forward spread more accurately identifies the signal for recession on the maturity spectrum, than yields do. At a given maturity, the forward rate is a yardstick of the short term rates that markets expect at that horizon plus a term premium. They argued that the six-quarter ahead rate identifies investors' expectations of ex-ante monetary policy decisions more accurately than the yield spreads hinged upon the ten-year yield. Attributing this to the fact that the ten-year yield is an average of the forward rates spanning over ten years, it dilutes the signal in the path of the short-term forward rates correlated with fluctuations in the business cycle. In accordance with this reasoning, they constructed a probit model to forecast the probability of recession with near term spread and the conventional 10-year minus 3-month yield spread as explanatory variables, and found that the near-term forward spread crowds out the effect of other slope variables in the model. They concluded that a negative value of near-term spread may only forecast recessions as the negative values capture the investors' expectations of a contractionary economy, inducing the Fed to cut the policy rate.

---

[2] Term Premium on a 10 Year Zero Coupon Bond. (2020, September 01). Retrieved from
https://fred.stlouisfed.org/series/THREEFYTP10

Another channel that connects the slope of the yield curve to the ex-ante real activity is the investors' risk tolerance that determines the risk premium. Benzoni and Chyruk (2018) constructed a dynamic term-structure model (DTSM) to study this channel. They decomposed the nominal yield on a Treasury security $r$ of a given maturity into its components – expected real interest rates, expected inflation, and risk premia that investors require to compensate them for facing inflation and real interest rates risks inherent in a security.

$r = E[\pi] + E[r] + IRP + RRRP + \varepsilon$, where:

$E[\pi]$ = expected trajectory of the consumer inflation rate over the treasury security's duration

$E[r]$ = expected trajectory of the inflation-adjusted real interest rates.

Together the sum: $E[\pi] + E[r]$ represents the expected path of the nominal interest rates

$IRP$ = inflation risk premium, $RRRP$ = real rate risk premium in the treasury yield, and

$\varepsilon$ : the actual treasury yield may deviate from the yield implied by the DTSM model by the error term.

Furthermore, they decomposed the yield spread i.e. the yield curve's slope, into the slope of the components: risk premium and expectations:

$slope\ r = slope\ E[\pi] + slope\ E[r] + slope\ IRP + slope\ RRRP$

Using data from 1985-Q1 to 2018-Q1, they examined the effects of employing these channels in the probit specifications to estimate the probability of a recession and the significant influences of each channel. More pronounced peaks before the inception of each of the three recessions in the sample suggest that their model's forecasts outperform those from the more traditional probit specifications.

Kelley and Benzoni (2018) modified the aforementioned decomposition to incorporate variables on a shorter time horizon. They found that whenever the Fed eases the monetary policy, noticed by either a lower real interest rate or a reduced expected real interest rate spread, the probability of a recession within a year rises. This contrasts with a diminished slope of risk premia is linked with either a lower or higher probability of recession, contingent on the origin of decline. More recently, a reduced slope of inflation risk-premium has signaled a greater chance of recession, and vice versa. Therefore, not all descent in the yield spread is a harbinger for the economy, and not all steepening are auspicious news either.

Merterns and Bauer (2018) constructed probit models to quantify the probability of recession at $t+12$ months based on the term at time $t$, contingent upon the term spread being above or below a particular threshold. The explanatory variables incorporated are term premium, natural interest rate, and the ratio of household net worth to income. Mertens and Bauer (2018) argue that quantitative easing had significantly depressed the term premium component of the long-term yields (Bonis, Ihrig, and Wei 2017). The yield curve flattens but doesn't raise recessionary risk, albeit the long-term yields decline, signifying that flattening may not always be bothersome. Keeping in mind that the term spread's predictive relationship does not reveal much about the causes of recession, we should note that yield curve inversions could cause recessions as elevated short term rates and tightening policies slow the economy. Alternatively, if investors expect a downturn, then the rising demand for safe, long-term Treasury bonds will reduce the long-term yields, inverting the yield curve.

Trubin and Estrella (2006) emphasized that the yield curve is a more forward-looking leading indicator as the recession signals that it produces are more advanced than those produced by other variables. Moreover, those signals are very sensitive to changes in the financial markets. As treasury securities don't face major credit risk premium, they are useful in forecasting the chances of a recession. They claim that using treasury yields whose maturities are far apart generate accurate results in forecasting the real activity. At the short-end of the curve, the three-month Treasury rate, when applied in conjunction with the 10-year Treasury rate yields robust and precise results. Furthermore, 10-year minus 2-year treasury rates graphically invert earlier and more frequently than those by 10-year minus 3-month Treasury rates spread, which is typically larger as depicted in figure 2. The evidence connotes that more pronounced inversions are correlated with deeper downturns.

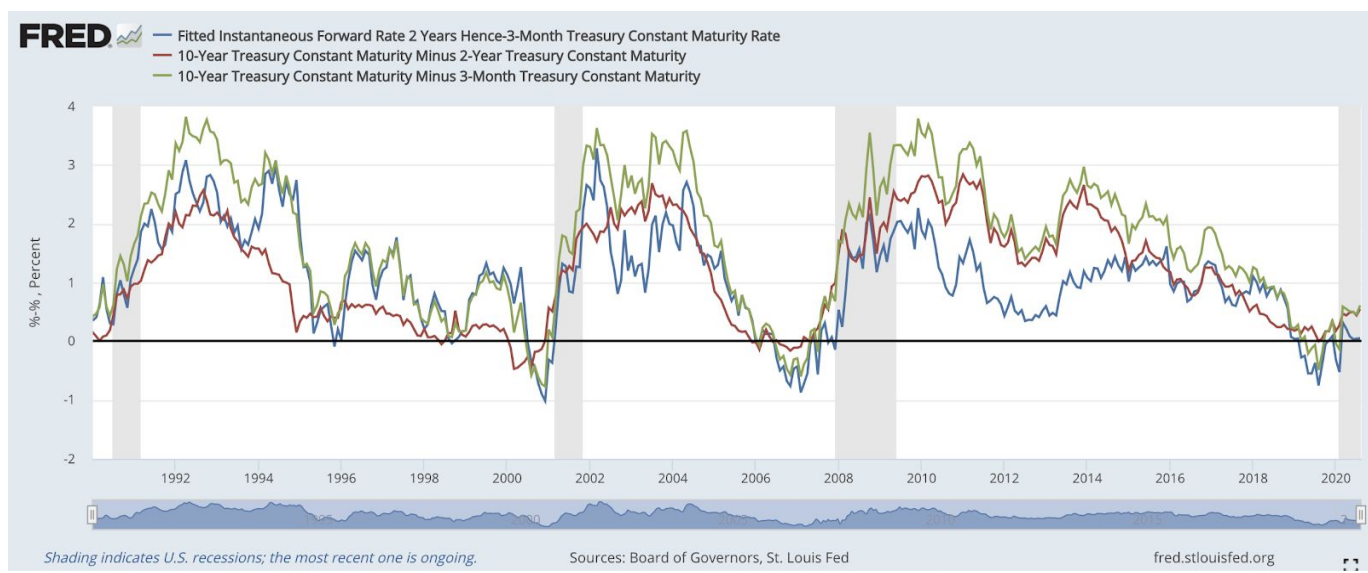

*Figure 2: Movements of different types of yield spread, including an example of the "near spread"*

Gogas (2015) analyzed the predictive power of the yield curve to forecast the US real GDP cycle during the global financial crisis through cross-validation. Applying a support vector machine (to classify a period into a recession or not), on data consisting of treasury bonds, bills of various maturities and real GDP from 1976-Q3 to 2011-Q4, they forecasted output and inflation gaps around the long-run trend. They defined a recession as the deviation of GDP under the long-term trend of output, and obtained the trend by decomposing the GDP into its cyclical and trend components using the Hodrick-Prescott filter. Transforming the cyclical component into an indicator variable, they forecasted the cyclical GDP in one, two and three-quarters ahead forecasting window. They estimated a linearly separable hyperplane coupled with a non-linear Kernel mapping procedure, by projecting the observations into a feature space (higher dimensional space), where the classes are linearly separable. Transforming using a linear kernel and radial basis function (RBF), generating a model using the latter resulted in the maximum out-of-sample accuracy of 66.7 percent and a cross validation test accuracy of 74.2 percent.

Bianchi et. al. (2018) examined the out-of-sample performance after modeling a diverse array of machine learning methods− ordinary least squares, partial least squares, random forests, randomized regression trees, penalized linear regressions and various neural networks− to forecast Treasury bonds across different maturities. Then, they contrasted the results with those obtained from principal component analysis (PCA), an unsupervised dimension reductionality method and concluded that machine learning techniques capture a vast proportion of variation in time series. Evidence suggests that the deep neural network significantly outperforms the forecasts calibrated from the alternative supervised learning and unsupervised method of PCA. Specifically, the increasing depth of neural networks (from shallow to three hidden layers) enhances the out-of-sample performance monotonically, particularly if the sample includes the ZLB. In that period, a pyramidal node structure of a neural network of four hidden layers produced by (Gu, Kelly and Xiu, 2018) outperforms the network with three layers. Moreover, incorporating economic priors about the role of variables boosts the network's performance. By clustering together hundreds of predictors consisting of macroeconomic data on bond return based on economic categories, and training a shallow network within each cluster or group yields a performance at par with a deeper neural network devoid of economic priors. Known as "group ensemble," economic priors on the network's architecture substantially affects the network's output. Overall, their results point the neural network's success to its property of capturing convoluted non-linearities embedded in the data and non-linearities within categories such as labor market, output, etc.

Das and Sambasivan (2017) constructed a Gaussian Process (GP) to model the yield curve data spanning from February 2006 to February 2017, and update the hyperparameters of GP as the algorithm received real-time data on the yield curve. They used a rolling window to train and test the performance of various methods and scrutinized the models' performance over three time durations−short-term: includes bond yields with upto 1 year in maturity, medium term: includes bond yields that mature in 2, 3 and 5 years; and long term: bonds that mature in 7, 10, 30 and 30 years. In forecasting the yields at the short-term region of the curve, their multivariate time series method performed well, whereas

GP produced more accurate results in the long and medium term regions of the yield curve. Obtaining higher accuracy in long term structures is harder than with short term structures as the observations in the long term region of the yield curve are far apart from each other than those in the shorter end of the yield curve. In juxtaposition to other methods, GP shows the probability and uncertainty estimates. Relatively, the Nelson Siegal model performs poorly in contrast with the dynamic GP and the multivariate time series approach.

Lu et. al (2019) developed a Long Short Term Memory (LSTM) model with an out-of-sample accuracy of 88.9 percent in forecasting the next day change in the credit spread. Comparing the results with those from other models− Vasicek model, Bayesian Additive Regression Tree (BART) and Random Forests, the BART model predicted the degree of credit spread quite well, but floundered in precisely forecasting the direction. Using the three day lagged data on credit spread in the LSTM model along with the feature variables, this model yielded the best ex-ante values of credit spread change. The advantage of LSTM is that it was able to gradually evolve and forget older data, while simultaneously learning patterns from recent history to extrapolate regime changes.

Niranjan et. al (2018) built two models−multilayer perceptron (MLP), and multivariate linear regression− to forecast the European yield curve at five forecasting horizons from time $t+1$ to $t+19$ days ahead. With a distinct set of features, they analyzed five variants of MLP such as a univariate model; a multivariate model with relevant explanatory variables including credit, commodities, equities, volatility and bond spreads; and two models with linear regression. Additionally, they employed two methods of multitasking learning−transformation into multiple task learning, and simultaneous modelling. The results from varying forecasting horizons attributed the best forecasting method to MLP with the relevant explanatory variables.

Torrent and Caldeira (2016) compared the forecast accuracy of a non-parametric functional data analysis (NP-FDA) model with parametric models in predicting the term structure of government bond yields. Amongst the methods they employ are: random walk, vector autoregression, autoregressive models and dynamic factor models such as those constructed by Diebold and Li (2006) and Nelson and Siegel (1987). Empirically, they show that the NP-FDA method furnished more accurate out-of-sample forecasts for all types of maturities and forecasting horizons than those generated by the dynamic Nelson–Siegel method of Diebold and Li (2006) and other aforementioned models.

## 3. Methods and Results[3]

**Principal Component Analysis (PCA)**

For PCA, I aggregated the data on the treasury yields − *tres*, of varying maturities from 3-month bills to 30-year bonds from October 10, 1993 to August 21, 2020. Figure 3 displays a three-dimensional plot of the data on yield curve. Whilst the considerable proportion of variation in the level is visually striking, the variation on the curvature and slope are less conspicuous.

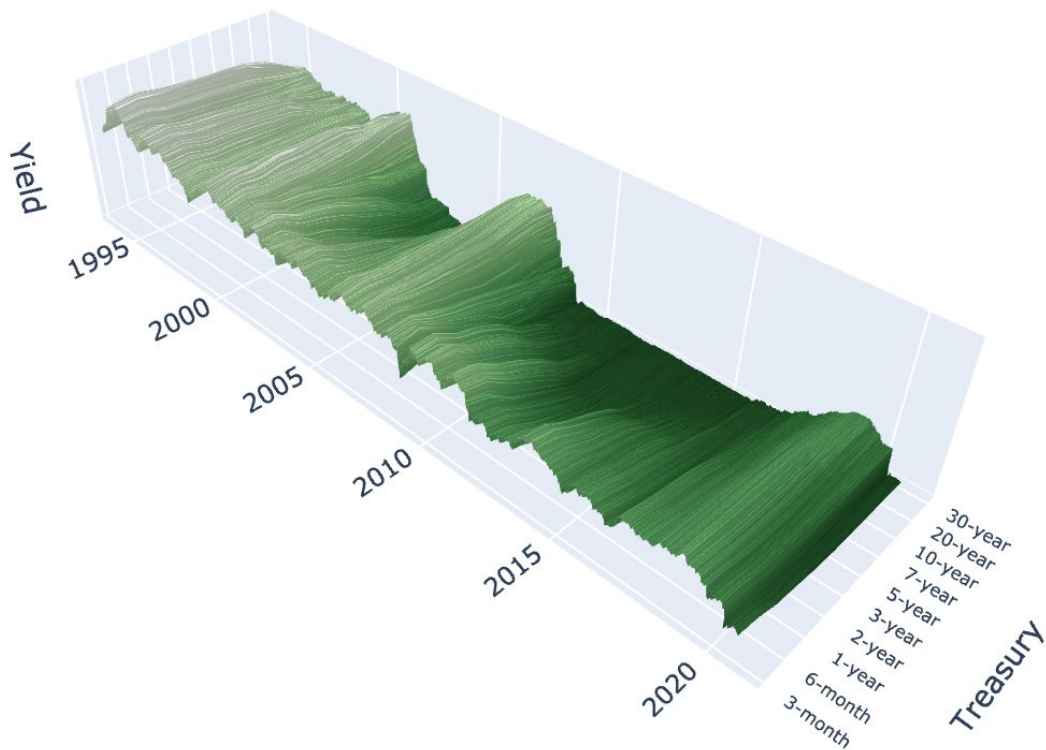

*Figure 3: 3-D display of the yield curve ranging from maturities : 3 months to 30-years Treasury yields.*

Then, I decomposed the yield curve via its principal components. The two features of any dataset include noise and signal and Principal Component Analysis (PCA) extracts the signal and diminishes the dataset's dimensionality. This is because it finds the fewest number of variables that explain the largest proportion of the data. It achieves this by transforming the data from a covariance or correlation matrix into a subspace (or an eigenspace) with fewer dimensions where all the explanatory variables are orthogonal to each other, thus avoiding multicollinearity and reducing noise. These vectors or axes of the eigenspace are known as the eigenvectors, and the eigenvalues determine the length of the vectors.

---

[3] Link to python (jupyter notebook) codes for replication: Github repository.

Whilst I calculated five principal components (PC) in total − $PC1, PC2, ..., PC5$ from the data on ten Treasury yields, I extracted the three latent factors that describe the dynamics of the yield curve − level, slope, and curvature. These are are measured by the first three principal components: PC 1, PC 2, and PC 3, as depicted in figure 4. The level refers to the parallel shifts in the yield curve; the slope is the changes in the short and long term rates evident by flattening and steeping of the curve, and twists denote the changes in the curvature of the model. After standardizing the data such that $tres \sim N(0, 1)$, I calculated the covariance matrix, and performed the eigendecomposition on the standardized data generated eigenvalues − $\lambda$ and eigenvectors − $v$. Whilst eigenvalues are the scalars of the linear transformation − $tres \times v = \lambda v$, eigenvectors are vectors whose direction remains unchanged even after applying the transformation. Then, I arranged $\lambda$ and $v$ based on decreasing $\lambda$, such that the first eigenvalue contributes the maximum variance to the $tres$ data. Finally, from the eigenvalues, I calculated the proportion of the total variance explained by each PC $i$

$$V(PC_i) = \frac{\lambda_i}{\lambda_1 + \lambda_2 + ... + \lambda_5}$$

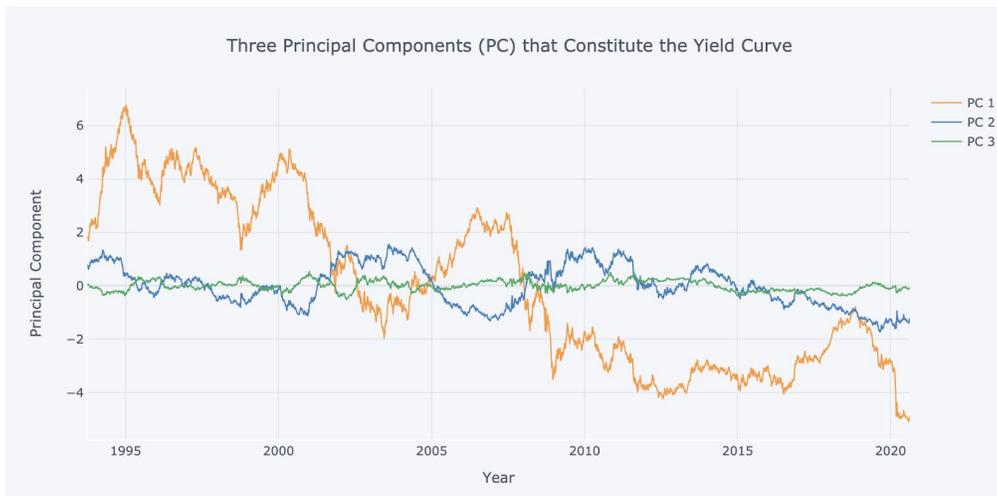

*Figure 4: The three PCs that form the basis of the Yield Curve*

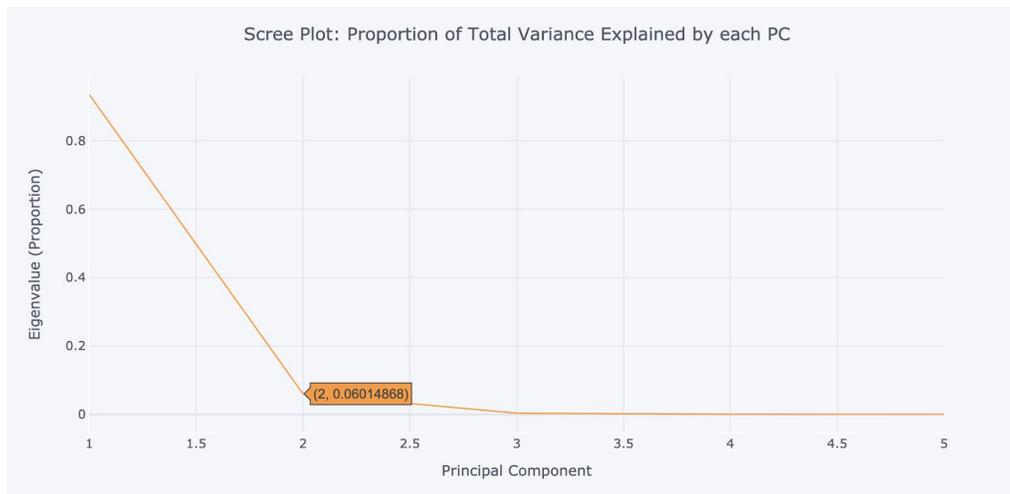

*Figure 5: Proportion of the total variance explained by each PC*

From figure 6, the plot of the first principal component looks very similar to the actual 10-year yield curve. This is consistent with our expectation as the first principal component explains 93.52% of the data as measured in table 1.

| Principal Component (PC) | Proportion explained by PC |
|---|---|
| 1 | 0.935264 |
| 2 | 0.060163 |
| 3 | 0.003582 |
| 4 | 0.000465 |
| 5 | 0.000249 |

Table 1. Proportion of the total variation explained by each PC

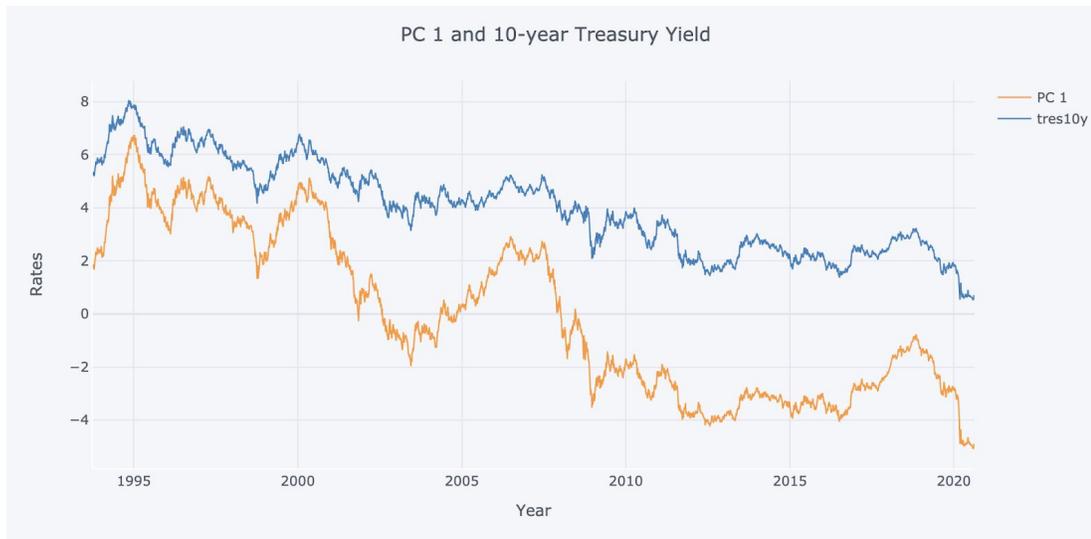

Figure 6: Similarities in the paths of the first PC and the 10-year Treasury bond

The second principal component depicts the slope, which in this case is the difference between the 10-year treasury bond and the 3-month treasury bill (10Y-3M spread), also called the yield spread. Visually, the slope appears nearly identical to PC 2 as shown in figure 7. Furthermore, the high correlation of 0.916 between the 10Y-3M spread $-yieldsp$, and PC 2 corroborates the evidence that the second principal component denotes the slope.

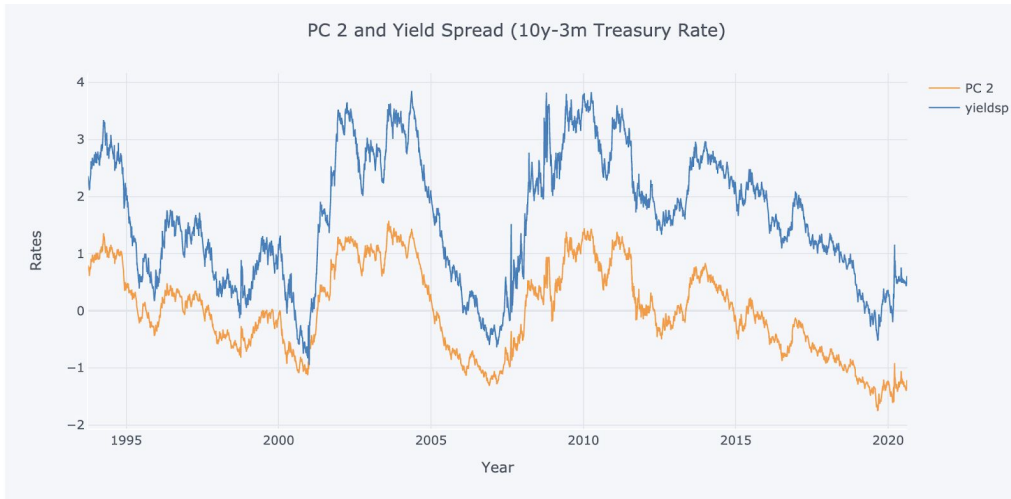

*Figure 7: Similarities in the paths of the second PC and the yield spread*

**Vasicek Interest Rate Model**

For univariate forecasting, the data comprises of *yieldsp* from January 4, 1982 to August 21, 2020. A stochastic technique of modeling the instantaneous movements in the term structure of forward interest rates is the Vasicek interest rate model. The factors that describe the trajectory of interest rates are time, market risk, and equilibrium value, wherein the model assumes that the rates revert to the mean of those factors as time passes. The larger the mean-reversion, the less is the probability that the interest rates will be closer to their current values, hence the rates will drift rapidly to their mean values over time. I have constructed a simple stochastic model to simulate the yield spread, which follows the same procedure as forecasting the short-term interest rates. This method employs maximum likelihood estimation to derive the parameters of the Vasicek model, which is of the form:

$dr_t = k(\theta - r_t)dt + \sigma dW_t$, where:

$k$ = strength of mean reversion or the speed at which the yield spread rates revert to the mean $\theta$.

$\theta$ = level of mean reversion or the long-run level of yield spread

$\sigma$ = volatility in yield spread at time $t$

$r_t$ = short rate (yield spread) at time $t$

$k(\theta - r_t)$ = expected changes in the yield spread or the drift term, also known as the mean reversion for Vasicek model.

$W$ = random market risk that follows a Wiener process

If the long-run mean value is less than the current rates, then the drift adjustment component will be negative. Consequently, the short term rate will be in close proximity to the mean-reverting level. If $r_t > \theta$, then the model pulls it down as $dr_t = k(\theta - r_t) < 0$, and if $r_t < \theta$, then it is pushed up as $dr_t = k(\theta - r_t) > 0$. Estimating the expected ex-ante yield spread rates, the Vasicek model calibrates the weighted average between the yield spread currently at time *t* and the expected long-term value $\theta$. In a nutshell, it forecasts the value of the yield spread at the end of a time period, contingent upon the recent volatility in the market, long-run average yield spread rate, and market risk factor. The concept of mean-reversion fails in periods of soaring inflation, economic stresses, and during crises. The interest rates (or the prices of securities) quickly incorporate the economic news when the mean-reversion parameter *k* is huge. In reverse, the effects will be longer if *k* is small. Using a closed form solution below that avoids "discretization errors"[4], I have simulated the paths of the yield spread:

$$r_{t_i} = r_{t_{i-1}} e^{(-k(t_i - t_{i-1}))} + \theta(1 - e^{(-k(t_i - t_{i-1}))}) + Z\sqrt{\frac{\sigma^2(1 - e^{-2k(t_i - t_{i-1})})}{2k}}, \text{ where } Z \sim N(0, 1)$$

I have employed the stochastic technique to model the yield spread ex-ante. Simulating from $r_0$ = *last observed value*, I have assumed that the long-run yield spread, $\theta = 1.75\%$, which is the mean of the data. Then, I generated a sequence of 10 ex-ante trajectories of the yield spread in each of the graphs below.

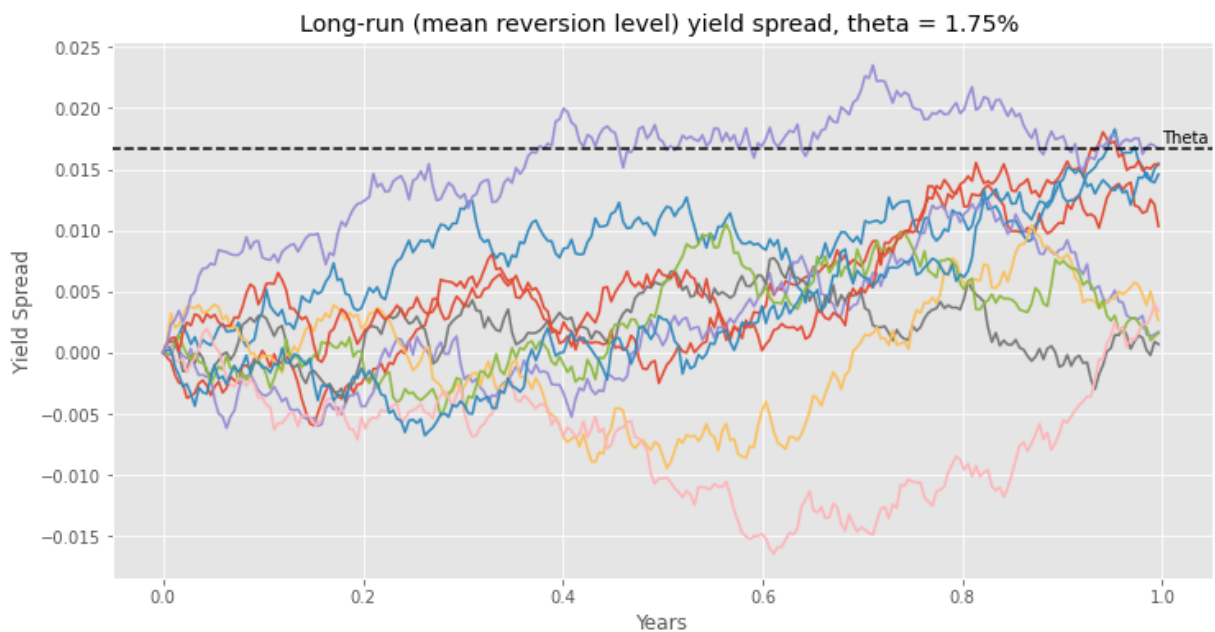

We observe the model's mean-reverting nature by specifying $r_0$ further away from $\theta$. Over time this pulls the yield spread towards $\theta$, the magnitude of *k* controls the speed of the reversion. As *k* grows, mean reversion quickens. Likewise, larger $\sigma$ widens the volatility and the potential distribution rate.

---

[4] Simulation of the short rate in the Vasicek model in R. (May 1, 2014). Retrieved from
http://delta9hedge.blogspot.com/2013/05/simulation-of-vasicek-interest-rates-in.html

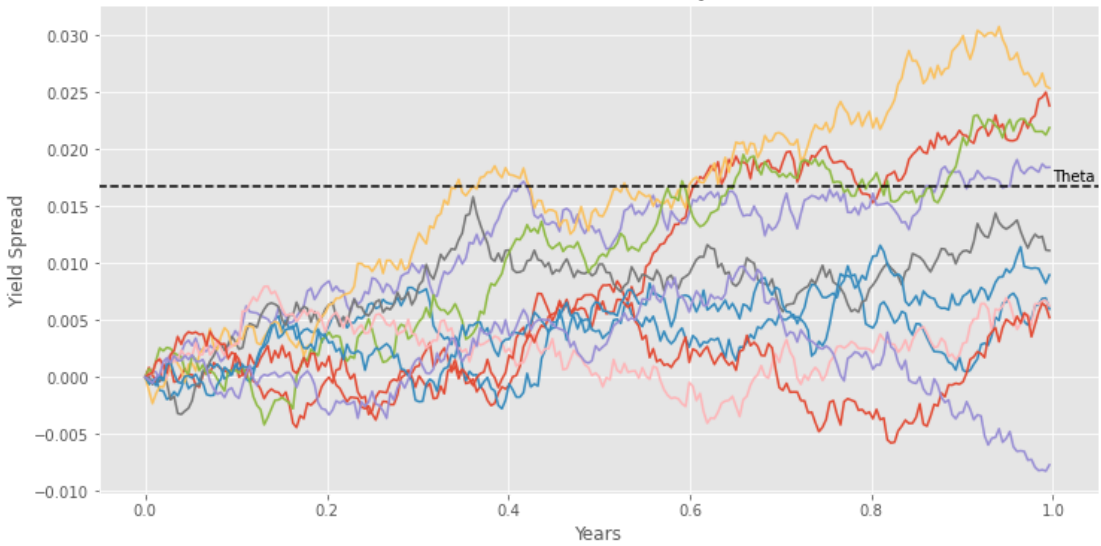

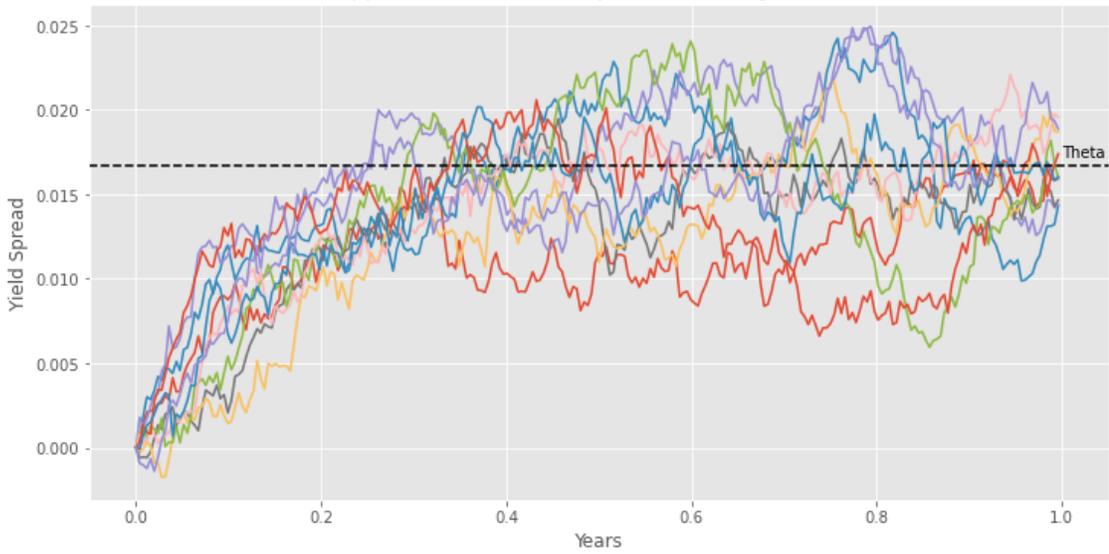

Increasing the speed of the mean reversion intensifies the pace at which *yieldsp* converges to its long-run level θ. When σ shoots up by 5 times, volatility rises, increasing the fluctuations in the ex-ante paths of *yieldsp,* making it harder to converge to the θ level of yield spread.

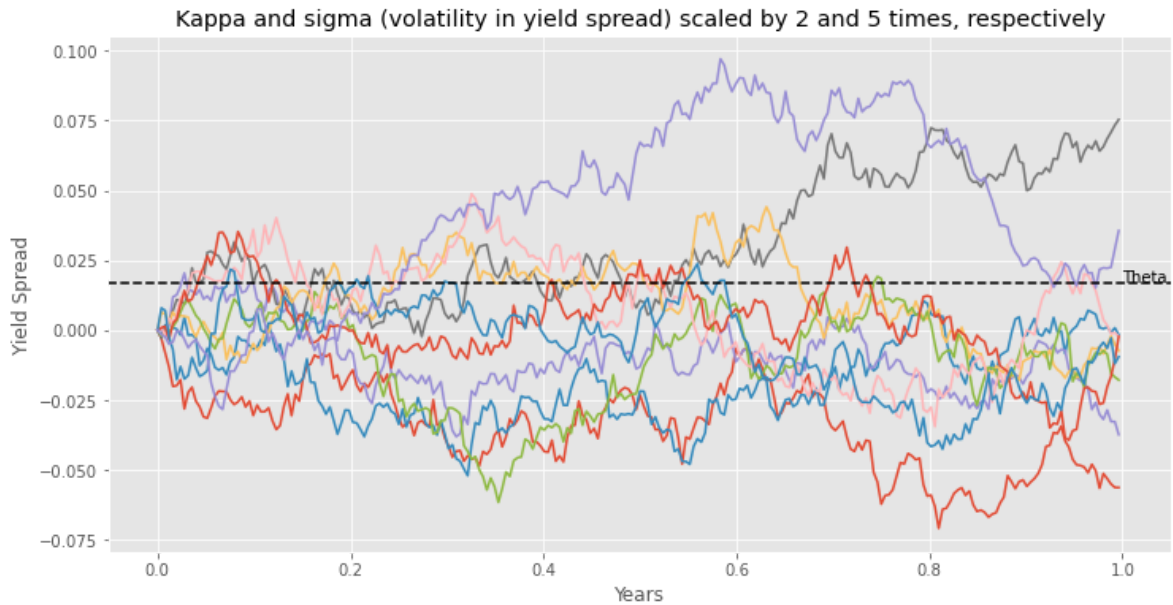

*Figure 8: 10 simulated trajectories of yieldsp under different parameter values of the Vasicek model*

Thus, knowing the long-run value of the short-term yield spread rates θ and the mean-reversion adjustment rate *k* enables us to calculate the evolution of the yield spread rates using the Vasicek model. Some of the caveats of the Vasicek model include that, firstly, the equation can only analyze one market risk factor at a time. Secondly, the long term rates have a relatively larger effect on the short-term rates than the short-term rates themselves. Lastly, the model overstates the long term volatility and understates the short term volatility. Next, I have applied the classical time series methods in forecasting the yield spread.

**Stationarity and ARIMA**

Before constructing a model, I checked for stationarity of the yield spread using the Augmented Dickey Fuller Test. Here, the null hypothesis is that a series has unit root or is non-stationary. A low p-value of 0.008 indicates that the data is stationary; resulting in the data having constant variance and covariance. Thus, the variance of *yieldsp* is not a function of time $t - V(yieldsp) \neq f(t)$, and $cov(yieldsp_t, yieldsp_{t+k}) = f(k) \neq g(t)$.

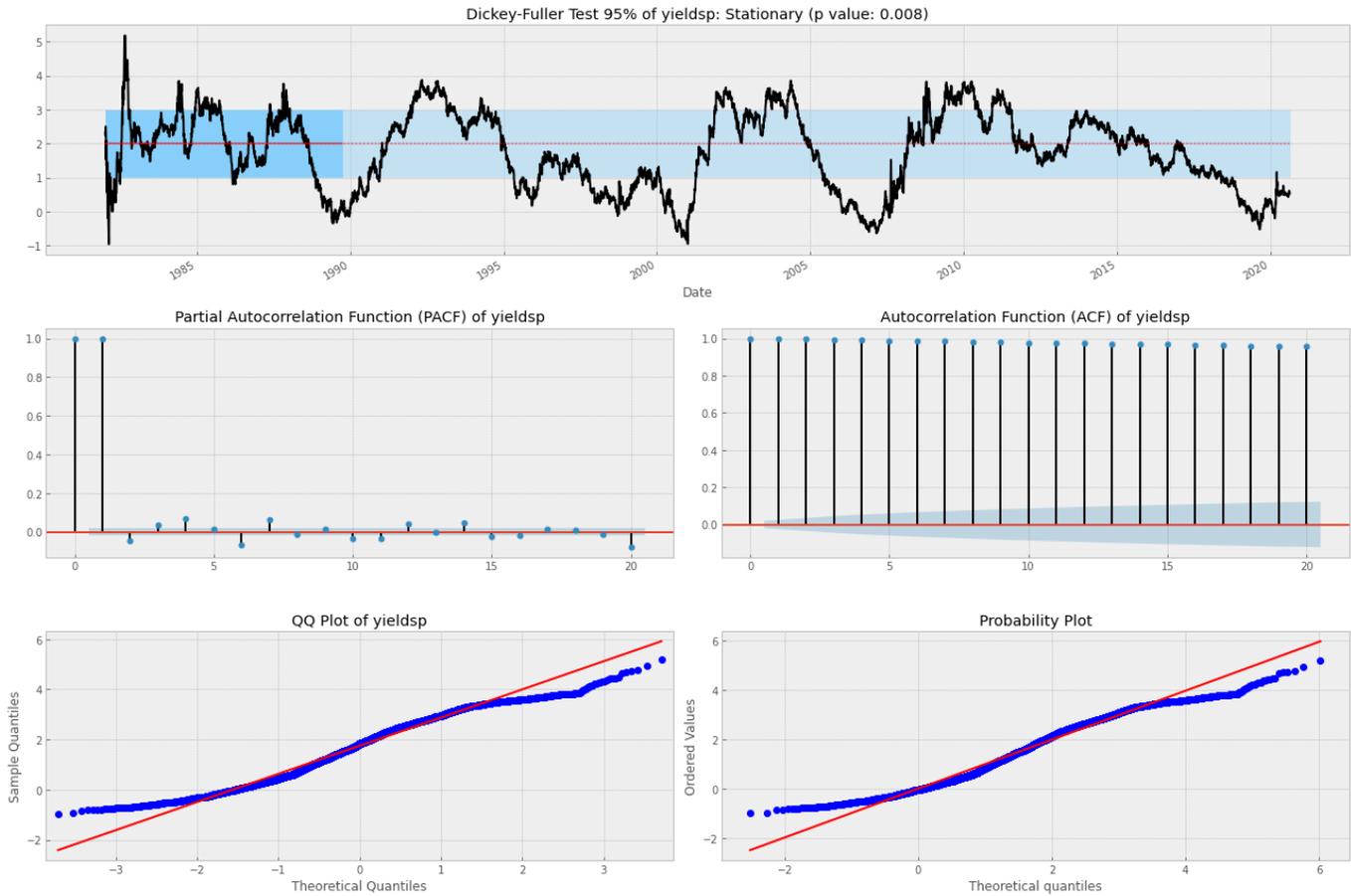

*Figure 9. Normally distributed, autocorrelated and stationary yieldsp*

The ACF plot in figure 9 shows a strong autocorrelation of lags as spikes very gradually reduce. The partial autocorrelation (PAC) measures the correlation between $y_t$ and $y_{t-n}$ after we control for correlations at intermediate lags. So, when we regress $y_t$ against a constant, $y_{t-1}, ..., y_{t-n}$, the PACF at lag $n$ is the regression coefficient on $y_{t-n}$. The PACF shows a significant lag for perhaps 3 days, with significant lags spotty out to perhaps 20 days. Due to considerable significant spikes shown in the ACF plot, I differenced *yieldsp* as *yieldsp_diff*, (written as $\Delta yieldsp$ in equations) and noticed that the series is not heavily serially correlated as previously depicted in the leveled data. From figure 10, the significant spikes of up to three lags in the ACF and PACF plots of the difference yield curve suggest that we can try $AR(3)$ and $MA(3)$ in the ARIMA model. First, I have modeled $ARIMA(1, 0, 3)$ on the leveled *yieldsp,* then developed $ARIMA(3, 1, 3)$ from difference $yieldsp - \Delta yieldsp$.

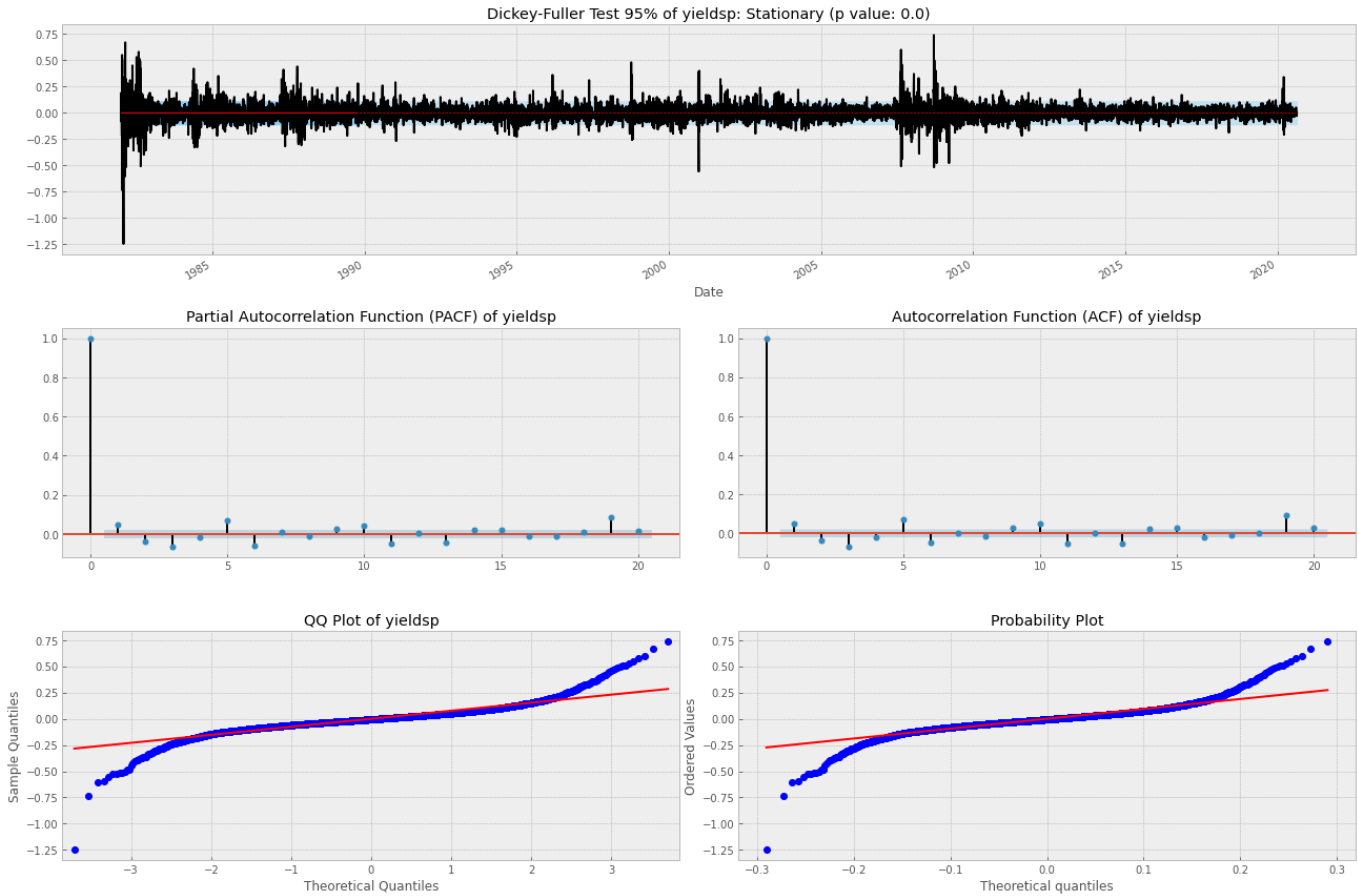

*Figure 10: Stationary, serially uncorrelated yieldsp but with slight heavy tails in the QQ-plot*

A time-series rests on the foundational assumption that the disturbance or the error term is a white noise process i.e. $E[\varepsilon_t] = 0$, $corr(\varepsilon_t, \varepsilon_{t-k}) = 0$ and $V[\varepsilon_t] = \sigma^2 = constant$. Thus, firstly we cannot use the last day's error term to forecast the current error term. Secondly, the error has a constant variance i.e. homoskedastic. Thirdly, the errors are serially uncorrelated. Alternatively, moving average (MA) models are extended versions of the white noise series which comprises the past forecast errors (unobserved white noise shocks $\varepsilon$) in a regression-like model. We use maximum likelihood estimation to fit in the parameters $-p, d, q$ of $ARIMA(p, d, q)$ and choose the model with the lowest Akaike information criterion (AIC). ARIMA is not sensitive to the data containing two types of stationarity: unit roots and hidden trends - seasonal, polynomial, linear, etc. While differencing eliminates any kind of polynomial trend, we may have to difference a multiple numbers of times to stationarize a series which has a higher degree polynomial trend. Fitting the $ARIMA(1, 0, 3)$ model yields in the entire dataset yields the following equation with $AIC = -20,952.034$:

$$yieldsp_t = 1.8092 + 0.9977\, yieldsp_{t-1} + 0.044\varepsilon_{t-1} - 0.0215\,\varepsilon_{t-2} - 0.0696\,\varepsilon_{t-3}$$
$$\quad\quad\quad (0.318) \quad\quad (0.001) \quad\quad\quad (0.010) \quad\quad\quad (0.011) \quad\quad\quad (0.011)$$

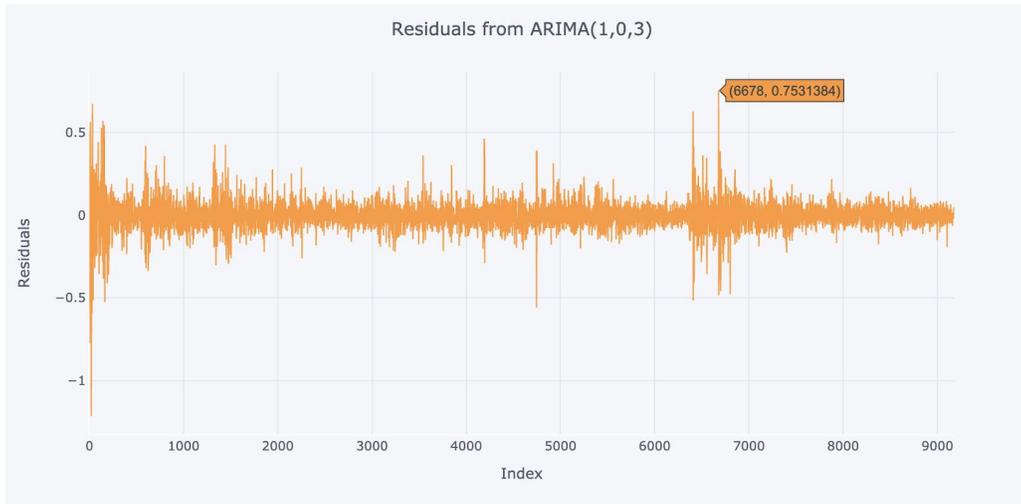

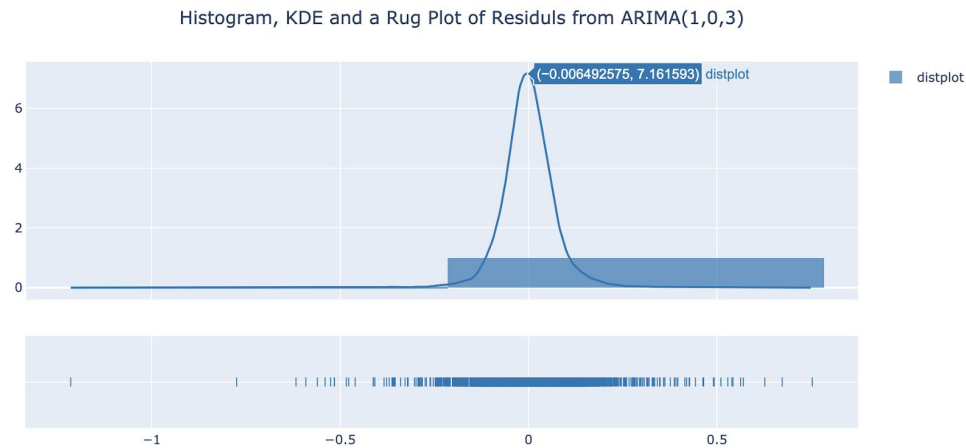

*Figure 11: (Top) Residuals from ARIMA(3,0,3) appear to be a white noise process, which is serially uncorrelated; (Bottom) Histogram of residuals are normally distributed*

Figure 11 and estimates from table 2 show the kernel density estimate (KDE) of the standardized residuals, which suggests the errors are Gaussian with a mean-centered at approximately zero, $-0.000062$, indicating no bias in the predictions. If the mean is non-zero, then the prediction may be biased positively. The line plot of residuals suggests that the model captures the trend information.

Next, I have fit $ARIMA(1,0,3)$ model on the train set and predicted for each observation in the test set. Evaluating using a walk-forward validation scheme, the model predicts for the next day $-t+1$, and then inputs the actual value of the yield spread for the next day to forecast the yield spread for the subsequent day $-t+2$, and so on. This procedure of rolling forecast recreates the ARIMA forecast each time the model receives a new observation. The test RMSE from the leveled model $-ARIMA(1,0,3)$ is $0.05185$. The ARIMA with the difference yield spread as shown below generates yields $AIC = -20,917.711$.

$$\Delta yieldsp_t = -3.341 \times 10^{-8} + 0.219\,\Delta yieldsp_{t-1} + 0.152\,\Delta yieldsp_{t-2} - 0.053\,\Delta yieldsp_{t-3}$$
$$(2.96 \times 10^{-7}) \quad\quad (0.005) \quad\quad\quad (0.068) \quad\quad\quad\quad (0.011)$$

$$-1.181\, \varepsilon_{t-1} - 0.0094\, \varepsilon_{t-2} + 0.1909\, \varepsilon_{t-3}$$
$$(-0.0094) \quad (0.1909)$$

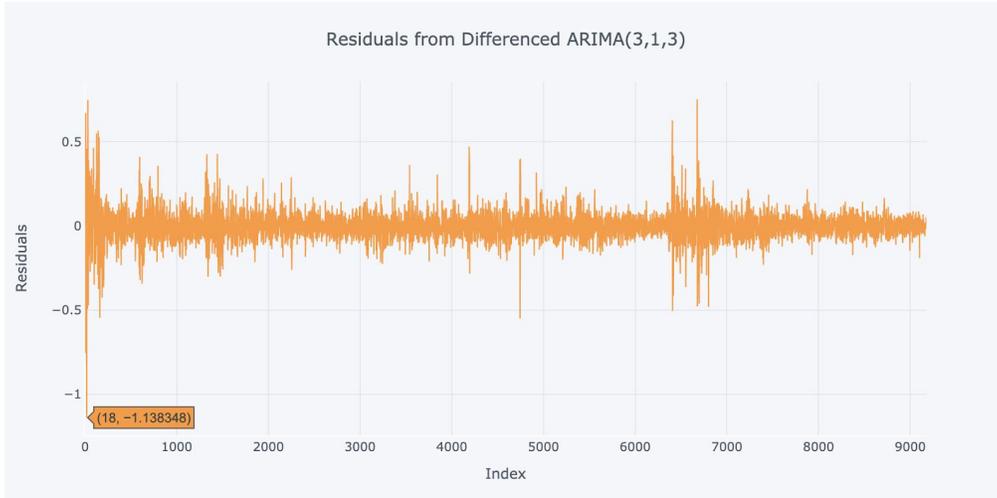

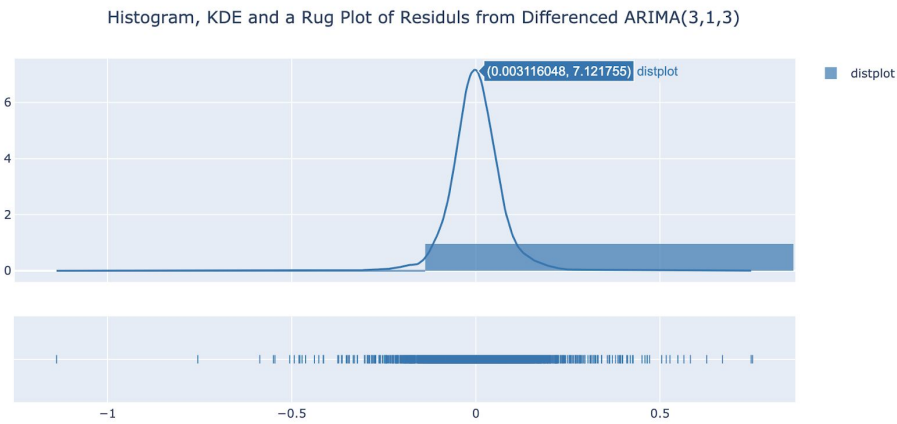

*Figure 12: (Top) Serially uncorrelated errors from the difference ARIMA; (Bottom) Histogram of those residuals are normally distributed*

|  | Residuals from $ARIMA(1, 0, 3)$ | Residuals from Difference $ARIMA(3, 1, 3)$ |
|---|---|---|
| Mean | $-0.000062$ | $0.000516$ |
| Standard Deviation | $0.077401$ | $0.077679$ |
| Minimum | − 1.214818 | − 1.138348 |
| 25% | − 0.037733 | − 0.037764 |
| 50% | − 0.001396 | − 0.000814 |
| 75% | 0.036517 | 0.037145 |
| Maximum | 0.753138 | 0.751493 |

Table 2. Descriptive statistics of the residuals from both ARIMA models

The residuals $\varepsilon_t$ from both the levelled and difference ARIMA models are stationary as the joint probability distribution of $\{\varepsilon_t, \varepsilon_{t+2}, ..., \varepsilon_{t+k}\}$ does not depend on $t$ for any $k$, and $E[\varepsilon_t \varepsilon_{t+k}] = 0$ for all $k \neq 0$. Following the same procedure of modeling the difference ARIMA in the train set using a walk-forward validation scheme as done previously (where the model is updated each time it received a new data), the test RMSE from the difference model − $ARIMA(3, 1, 3)$ is 0.05615. Figure 13 compares the ex-ante *yieldsp* from the leveled and the difference ARIMA models, and the ex-post values of *yieldsp* for the most recent period from April 4, 2020 to August 21, 2020.

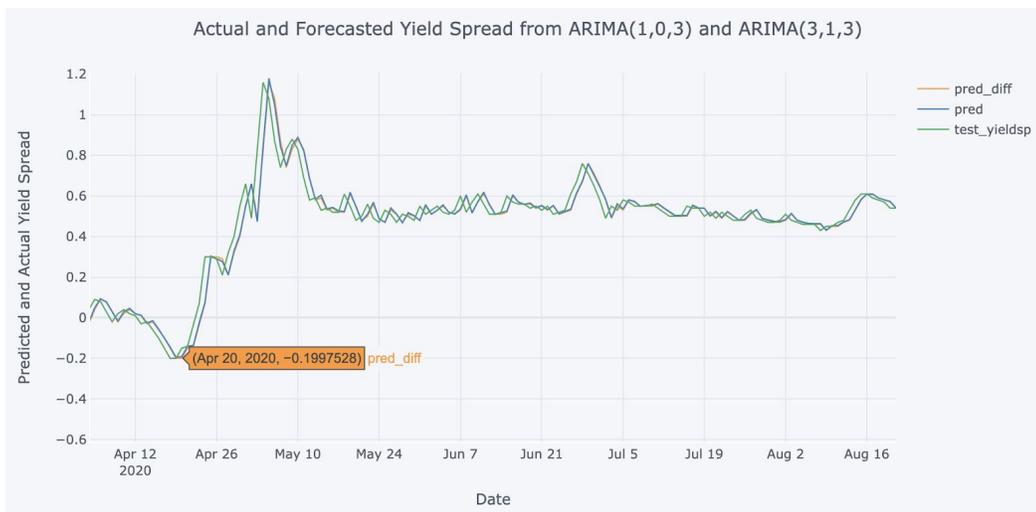

Figure 13: Actual vs. predicted yield spread in the test set from both levelled and difference ARIMA models

**Conditional Heteroskedasticity − (G)ARCH**

Mainy, the deficiencies of ARIMA models are − firstly, they are not conditionally heteroskedastic i.e. they don't account for volatility clustering. Secondly, the forecast width of ARIMA models is fixed as it linearly models the data. Hence, I used the Autoregressive Conditional Heteroskedastic (ARCH) model and Generalised Autoregressive Conditional

Heteroskedastic (GARCH) model to show heteroscedastic variances. In financial economics, a change in variance in one time may change the variance in the same direction in the subsequent time, as observed in the yield spread data. For instance, a drop in yield spread when recession is imminent worries investors about deteriorating economic and financial conditions, leading to further drop in yield spread. This can occur when the yields on long term treasuries plummet, shrinking the gap between short and long term yields, and may cause *yieldsp* to be negative when short term yields surpass the long term yields, signalling a contractionary economy. If the variance of the yield spread represents the riskiness of the spread, then certain time periods are riskier than others; connoting that magnitude of error in certain times is greater than in other times. Heightened riskiness depicts volatility clustering.

Moreover, as these risky times are not randomly scattered across the daily time series, a degree of autocorrelation is present in the yield spread. This downward trajectory of volatility is evidence of heteroskedasticity where errors are serially correlated. Thus, the series of errors is conditionally heteroskedastic. Even in heteroskedasticity, the OLS regression coefficients are unbiased, but the confidence intervals and standard errors generated conventionally appear very narrow, giving a misleading sense of precision. Rather than considering this as a problem to be resolved, we use (G)ARCH models that treat heteroskedasticity by modeling the variance. Resultantly, it not only resolves the defects of OLS, but also forecasts the variance of residual. The tricky aspect of conditional heteroskedasticity is that the ACF plots of a very volatile series may misleadingly appear to be a stationary discrete noise process, albeit the series actually has unit root with varying variance. To accomodate this conditional heteroskedasticity, I constructed a GARCH model as it entails autoregressive parameters of the variance − incorporating past changes in variances.

An extension of ARCH, GARCH encompasses an MA component alongside the usual AR component from ARCH to model not only the conditional change in variance but also adjustments in the time-dependent variance over time. For instance: conditional rise and fall in variance. Also, it includes the lagged variance terms and lag residual errors derived from a mean process. The parameter *p* in GARCH refers to the number of lag variance terms, whereas *q* refers to the number of lag residual errors included in the GARCH model. This culminates to $GARCH(p, q)$. Subsuming ARCH models, $GARCH(0, q)$ is equivalent to an $ARCH(q)$ model. Hence, $p = 0$, implies an $ARCH(q)$ process. If both $p = q = 0$, then the series is a white noise.

In $GARCH(p, q)$, $\varepsilon_t = \sigma_t w_t$, where $w_t \sim N(0, 1)$

$\varepsilon_t$ is a generalized autoregressive conditional heteroskedastic model of order *p* and *q* if:

$$\sigma^2_t = \alpha_o + \sum_{i=1}^{p} \alpha_i \sigma^2_{t-i} + \sum_{j=1}^{q} \beta_j \varepsilon^2_{t-j}$$

Useful for forecasting volatility, (G)ARCH models the change in variance over a time period as a function of the residual errors from a mean process. These models explicitly differentiate between conditional and unconditional variance, and let the former change as a function of historical errors. I specified a lag parameter *q* to state the number of historical residual errors incorporated in the model. (G)ARCH are applied to stationary series - devoid of seasonal and trend components, but can have non-constant variance. These models project the ex-ante variance for a given time horizon. (G)ARCH has a mean of 0, uncorrelated processes with heteroskedasticity conditional on the lagged values, but constant unconditional variances. Empirically, after constructing ARIMA, we can use (G)ARCH to model the expected variance on the residuals, provided that the residuals are not serially correlated and don't have any trend and seasonal patterns.

When fitting an $AR(p)$, figure 14 demonstrates the decay of the $p$ lag displayed in the ACF plot of the residuals and the squared residuals. To apply GARCH we need to ensure that the mean of the residuals is 0, and from the descriptive statistics in table, the mean of the residuals obtained from $ARIMA(1,0,3)$ is $-0.000062$, which is very miniscule and approximately 0.

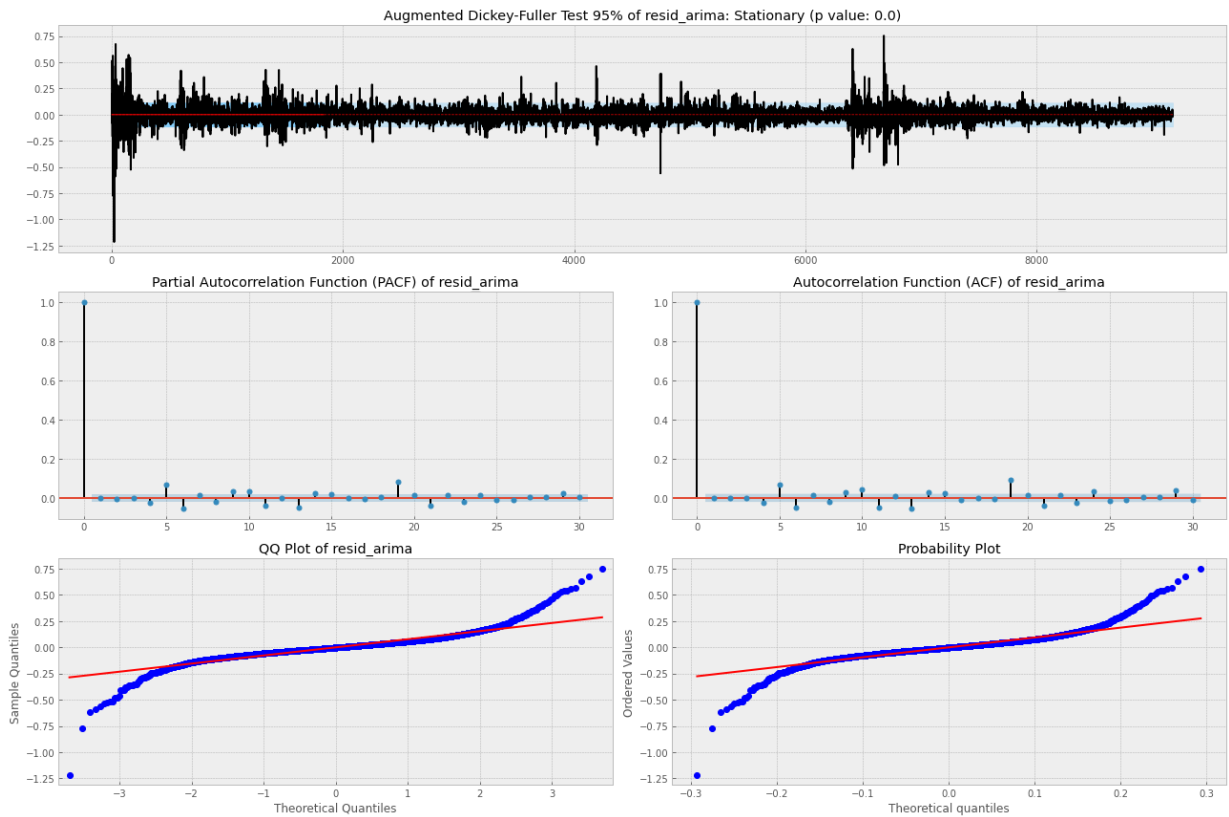

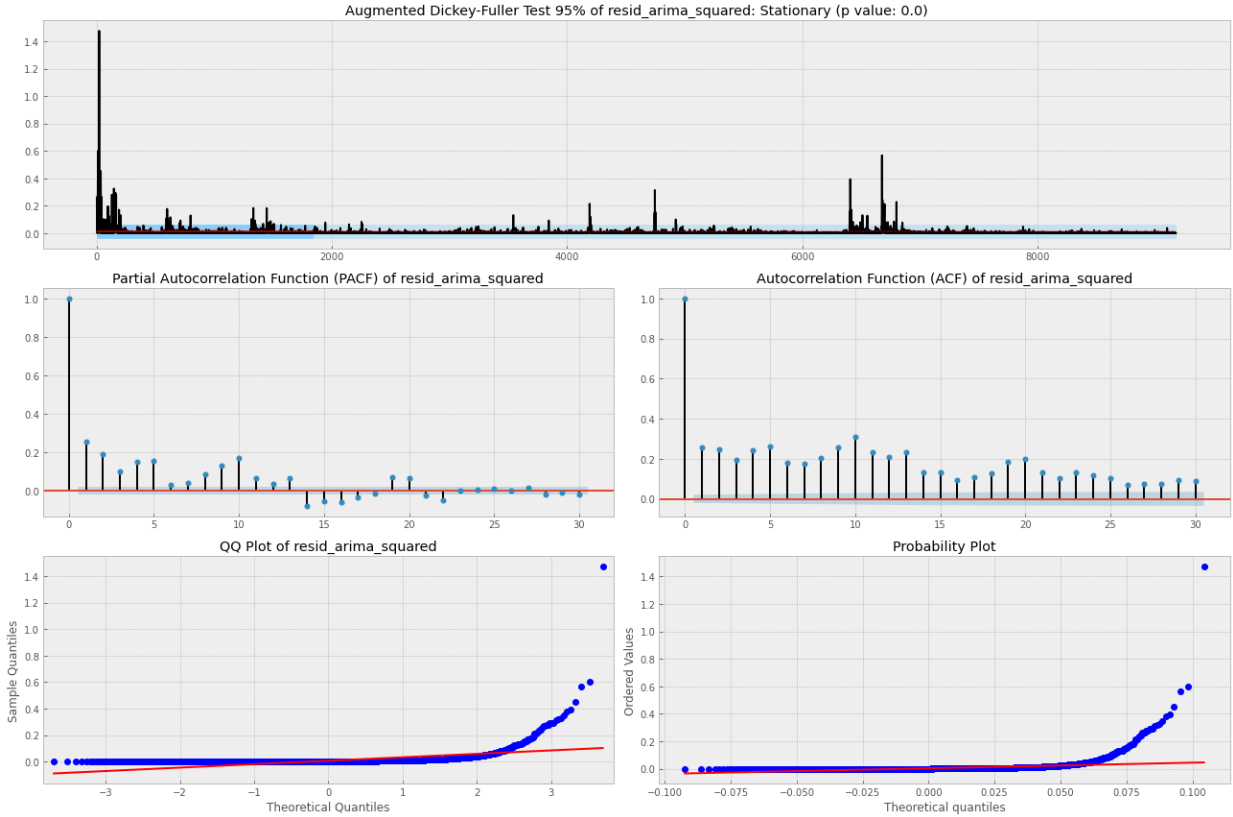

*Figure 14: Test for stationary, normality and autocorrelation of the residuals and squared residuals from ARIMA(1,0,3)*

Whilst the residuals from $ARIMA(1,0,3)$ appear to be a white noise process, the squared residuals are highly serially correlated. The slow decay of successive lags indicates conditional heteroskedasticity. The Q-Q plot indicates that while the points fall along the line in most parts of the graph, they curve-off in the extreme right hand side, suggesting that the residuals are more extremely valued than otherwise found in a normally distributed data. Given that the lags in both the PACF and ACF plots are significant, the model would be better fit with both AR and MA parameters. So, I have fit a $GARCH(1,3)$ model and checked how the residuals from it behave. From the generalized $GARCH(p,q)$, we can write $GARCH(1,3)$ as:

$$\sigma^2_t = \alpha_o + \alpha_1 \sigma^2_{t-1} + \sum_{j=1}^{3} \beta_j \varepsilon^2_{t-j} \Rightarrow \sigma^2_t = \alpha_o + \alpha_1 \sigma^2_{t-1} + \beta_1 \varepsilon^2_{t-1} + \beta_2 \varepsilon^2_{t-2} + \beta_3 \varepsilon^2_{t-3}$$

$$\sigma^2_t = 1.7618 \times 10^{-3} + 0.1145\, \sigma^2_{t-1} + 0.8015\, \varepsilon^2_{t-1} + 2.1119 \times 10^{-11} \varepsilon^2_{t-2} + 0.0840\, \varepsilon^2_{t-3}$$

$\qquad\quad\; (3.882e-04) \qquad\;\; (1.359e-02) \qquad\quad (0.330) \qquad\qquad\;\; (0.597) \qquad\qquad\;\; (0.281)$

The estimates of all parameters except $\beta_3$ fall within their respective confidence intervals, and AIC is $-24,666.6$. While the residuals (from GARCH) plot (not shown) looks like a realization of a discrete white noise process, indicating a good fit, the squared residuals are still not fully white noise, albeit less serially correlated than the squared residuals from $ARIMA(1,0,3)$. So, $GARCH(1,3)$ has not properly "explained" the serial correlation present in the squared residuals, inhibiting me from predicting in the test set.

**Multivariate Forecasting**

Next, I have forecasted *yieldsp* using multivariate daily data from January 1,1990 to June 1, 2020 obtained from FRED. The variables incorporated are :

1. Real-time Sahm Rule recession indicator − *sahm* : It signals the beginning of a recession when the three month moving average of the U3 unemployment rate increases by at least 0.5 percent in contrast to its lowest level in the previous twelve months. It is based on real − time data i.e. the values of the current and historical unemployment rate available at a particular month. Rising percentage points of *sahm* from its average are indicative of impending contraction, and the number accelerates during a recession.
2. Fitted instantaneous forward rate 1-year rate hence − *forward*1*yr* : generated by Kim and Wright (2005) when they fit a "three-factor arbitrage-free term structure model." It extracts the markets' expectations of the ex-ante paths of variables such as the short-term rates, term premium in bond yields, etc. A negative correlation with yield spread shows that it is inversely related with *yieldsp.*
3. NBER based recession indicator − *rec_ind* : $1\{recession\}$. The dummy variable of 1 indicates a recessionary or a contractionary period, whereas 0 indicates an expansionary period in the US business cycle. *yieldsp* enters the negative territory for a month as a harbinger of a culminating recession, switching the values in the dataset from 0 to 1.
4. Term premium on a 10-year zero coupon bond − *termpr* : like *forward*1*yr*, Kim and Wright (2005) constructed it using fitting the arbitrage free term-structure model on the US Treasury yields. Departing from the expectations hypothesis, the term premium is the difference between the yield and the average expected short rate over the life of the coupon bond. We can attribute the recent declining trend in the term premium due to several factors such as stable below-target inflation rate, more effective and explicit forward guidance, and quantitative easing during the Zero Lower Bound. Absence of inflation and other monetary or fiscal shocks may have diminished compensation associated with lower risk. A higher term premium due to systemic risks can arise when the yield spread is narrow or on the verge of being inverted.
5. University of Michigan Inflation Expectations − *infexp* : From the Survey of Consumers, it reflects the latest changes in the prices that consumers expect in the next one year. In theory, inflation expectations rise when investors expect the economy to heat during easing monetary policy, and vice-versa. This occurs when the short-end of the yield curve is at a very low level while the long-term yields may be high, widening the yield spread.
6. CBOE Volatility Index − *vix* : conveys investors' expectations of the short term volatility embedded in the prices of stock index options. Higher volatility usually occurs during unstable financial markets, typically arising during or before a recession when the yield spread significantly narrows and may turn negative.
7. TED spread − *ted* : It is the difference between the US dollar value of 3-month LIBOR rate and the 3-month Treasury bill.

8. 1-year Treasury constant maturity minus the federal funds rate − $1yrffr$ : On average this alternative measure of yield spread is much lower (0.1126) than the 10Y-3M measure of spread (1.7102) as the long end of the yield curve is only 1 year as opposed to 10 years.

As opposed to univariate forecasting, the range of *yieldsp* in multivariate modeling is smaller, and the Augmented Dickey-Fuller test from table 3 suggests that the variable is no longer stationary. Furthermore, to check if the sample data on the yield spread arises from a Gaussian distributed population, I have used the goodness-of-fit measure : D'Agostino's $K^2$ Test. Transforming the sample skewness and kurtosis derives the test statistic. The null hypothesis is that the data is normally distributed, and the alternate hypothesis is that the data is kurtic and/or skewed (not normally distributed)

The small p-value of 0.00 rejects the null hypothesis. Hence, the sample of yield spread is not normally distributed.

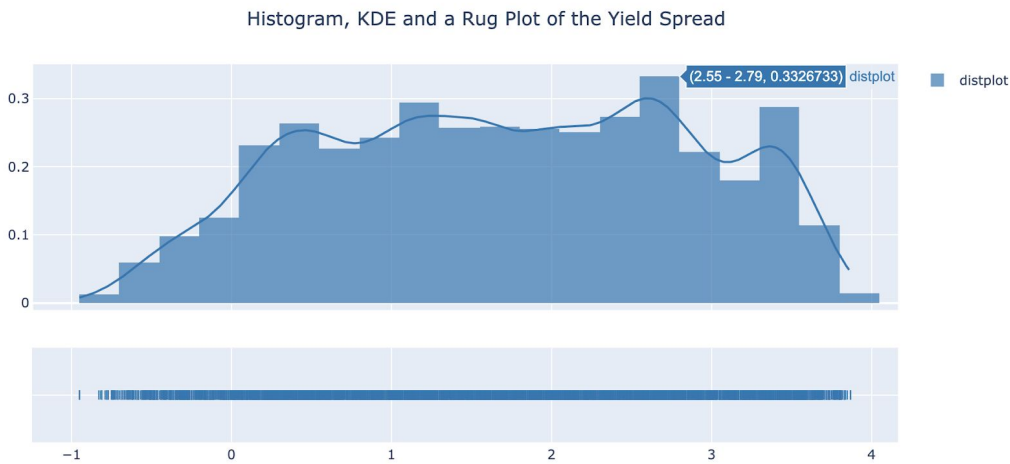

Figure 15: Heavy tails and moderately skewed yieldsp

Describing the sharpness of the peak of the normal distribution, the kurtosis is close to 0. The distribution has heavier tails if kurtosis is less than 0, and lighter tails if it is greater than 0. The distribution is symmetrical if the kurtosis lies in the range of (− 0.5, 0.5). The data is moderately skewed if the skewness is within (− 1, − 0.5) or within (0.5, 1). For a highly skewed data, the skewness values are either less than − 1 or greater than 1. The yield spread data is moderately skewed as its skewness is − 0.0788 , and has heavy tails (very kurtotic) because the kurtosis is − 1.0258.

| Variable | p-value | Inference |
|---|---|---|
| *yieldsp* | 0.1305 | non-stationary |
| *ted* | 0.0000 | stationary |
| *1yrffr* | 0.0000 | stationary |
| *forward1yr* | 0.5194 | non-stationary |
| *rec_ind* | 0.0021 | stationary |
| *termpr* | 0.5938 | non-stationary |
| *sahm* | 0.2994 | non-stationary |
| *infexp* | 0.0000 | stationary |
| *vix* | 0.0000 | stationary |

*Table 3. Test for stationarity of yieldsp and the explanatory variables*

After first differencing the variables *yieldsp, forward1yr, termpr* and *sahm,* they become stationary.

**ARIMA With Exogenous Variables and Granger Causality**

The Seasonal ARIMAX equation consists of a univariate time series as a dependent variable, at least one exogenous variable, and may have their lags. We can forecast the future values of $yieldsp_{t+h}$ only if we have either the ex-ante or ex-post values of the exogenous variables. However, the lack of those forecasts has inhibited me from forecasting $yieldsp_{t+h}$, i.e. the ex-ante *yieldsp* beyond those in the test. On the training set, I have modeled $SARIMAX(2,1,2)$ without any seasonal component as *yieldsp* is not seasonal, and incorporated the stationary exogenous variables such as *ted*, $\Delta termpr$, $\Delta forward1yr$, etc as additional variables to forecast *yieldsp*. Resulting in the AIC of $-44321.148$, $SARIMAX(2,1,2)$ is:

$yieldsp_t = 0.4153\ ted_t - 0.0051\ rec\_ind_t + 0.0055\ yrffr_t - 8.782 \times 10^{-5}\ vix_t + 2.4098\ \Delta termpr_t$

        (0.002)            (0.021)          (0.000)          (0.000)           (0.010)

$- 0.4141\ \Delta forward1yr_t - 0.0166\ infexp_t + 0.0351\ sahm_t + 0.0055\ Iyrffr_t + 1.0867\ yieldsp_{t-1}$

   (0.005)               (0.018)          (0.016)         (0.000)         (0.020)

$- 0.6744\ yieldsp_{t-2} - 1.0131\varepsilon_{t-1} + 0.2122\varepsilon_{t-2}$

(0.018)          (0.022)      (0.020)

The model diagnostics in figure 16 showcase if any of the OLS assumptions have been violated.

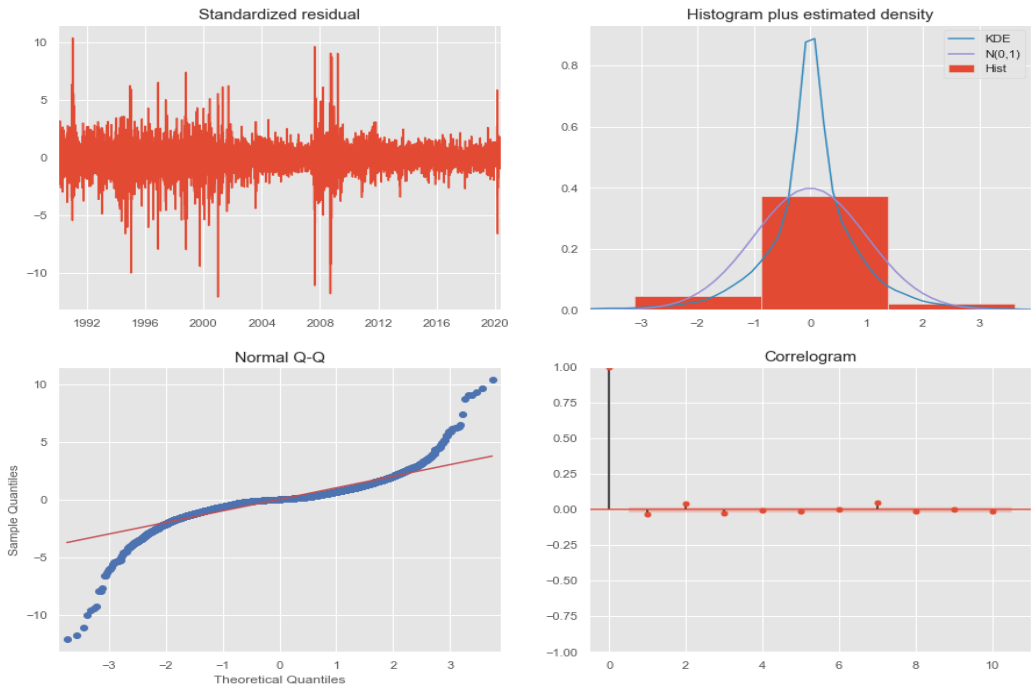

*Figure 16: Stationary and serially uncorrelated residuals*

Ensuring that the residuals of our model are uncorrelated and normally distributed with zero-mean, the model diagnostics imply that the residuals have gaussian distribution. The red KDE line closely follows the $N(0, 1)$ line, indicating that the residuals are normally distributed. The standardized residuals appear to be white noise and don't display seasonality, also evident from the correlogram - the time series of the residuals have very low serial correlation with its lagged values. However, the Quantile-Quantile (QQ) plot shows heavy tails as whilst the points lie along the line in the middle, they curve off in the extreme ends. This implies that the residuals have more outliers or extreme values than expected if the residuals were normally distributed. Using the test values of these exogenous variables, I predicted the test set *yieldsp* for the test set and the confidence interval associated with each forecast. At each observation, I have generated forecasts using the full history upto that point. The root mean squared error of the forecasts is 0.3458, more than those produced by univariate models. The monotonically increasing bands around the forecasts in figure 17 indicate that the model's beliefs about the distant future is less precise than those for the near future.

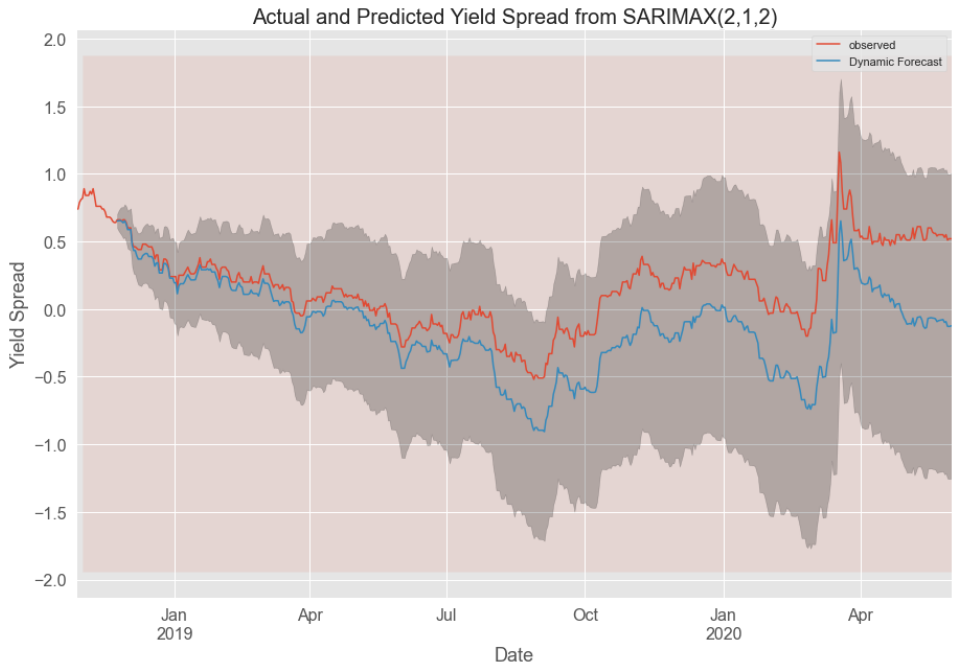

Figure 17: Actual and forecasted values of yieldsp from the SARIMAX model

Another forecasting method employed is the reduced-form Vector Autoregression (VAR) . But before constructing, I checked for the presence of Granger causality to determine which variables have causal relationship with $\Delta yieldsp,$ since the Granger Causality test is only applied to stationary variables. Here, the null hypothesis is that lagged values of an explanatory variable $x_{jt}$ in the matrix of explanatory variables $-X,$ do not explain the variations in $\Delta yieldsp_t.$ In that case, $x_{jt}$ doesn't granger cause $\Delta yieldsp_t.$ Specifying the maximum number of lags specified as 40, the equation to test the null hypothesis is:

$$\Delta yieldsp_t = \alpha_0 + \sum_{i=1}^{40} \alpha_i \, yieldsp_{t-i} + \sum_{j=1}^{40} \beta_j x_{t-j}, \tag{1}$$

where $x_j$ refers to a stationary explanatory variable from the matrix:

$X = \{\Delta\, termpr,\ \Delta forward1yr,\ ted,\ vix,\ rec\_ind,\ infexp,\ \Delta sahm,\ 1yrffr\}$

For instance, to calculate the granger causality of *ted* on *yieldsp,* the equation is:

$$\Delta yieldsp_t = \alpha_0 + \sum_{i=1}^{40} \alpha_i \, yieldsp_{t-i} + \sum_{j=1}^{40} \beta_j ted_{t-j}$$

While I calibrated the granger causality of every stationary variable on each other, table 4 below displays the results the p−values of the test, signifying the causality of the stationary explanatory variable on *yieldsp* in the train set.

|  | $\Delta yieldsp$ | Conclusion |
|---|---|---|
| $\Delta\ termpr$ | 0.000 | $\Delta\ termpr$ granger cause *yieldsp* |
| $\Delta\ forward1yr$ | 0.000 | $\Delta\ forward1yr$ granger cause *yieldsp* |
| *ted* | 0.000 | *ted* granger cause *yieldsp* |
| *vix* | 0.000 | *vix* granger cause *yieldsp* |
| *rec_ind* | 0.000 | *rec_ind* granger cause *yieldsp* |
| *infexp* | 0.1 | *infexp* don't granger cause *yieldsp* |
| $\Delta\ sahm$ | 0.0517 | $\Delta\ sahm$ granger cause *yieldsp* |
| *1yrffr* | 0.000 | *1yrffr* granger cause *yieldsp* |

*Table 4: Test for Granger Causality*

All the variables except *infexp* and $\Delta\ sahm$ Granger causes *yieldsp* as each of their p−values are less than the 5 percent significance level. The lagged values of all other variables in *X* are retained in the regression equation (1) since they are individually significant i.e. they add explanatory power to the regression according to the F− test. As *infexp* and $\Delta\ sahm$ don't granger cause $\Delta yieldsp$, their lagged values are not retained in equation (1). So, I have not included *infexp* and $\Delta\ sahm$ in the VAR discussed below.

**Vector Autoregression (VAR)**

Extending the idea of univariate regression to a multivariate time series, the reduced-form vector autoregression of order $l - VAR(l)$ forecasts *yieldsp* using the *l* lagged values of itself and the stationary variables. To approximate the process well in the absence of MA terms, we may require a larger AR order. Thus, more parameters will be estimated, increasing the fitted variance. To tame the variance, we can regularize (or apply a shrinkage term to the VAR model). Firstly, I determined the optimal VAR order from the lag order criteria $- AIC$. From a pool of models, AIC selects a model such that it generates the smallest one-step ahead squared forecast error. Setting the range of lags from 1 to 50, figure 18 displays the optimal lag order of 33, and the AIC from $VAR(33)$ is $- 40.8102$. Graphically, the AIC diminishes as the lag order increases:

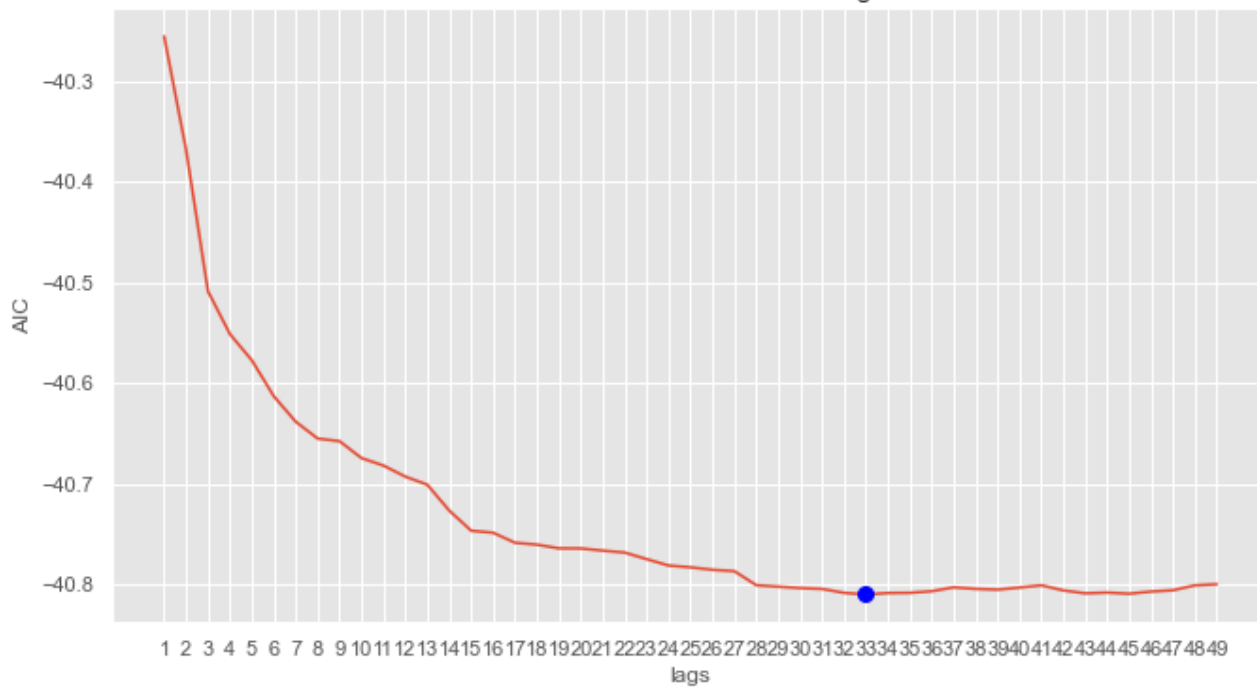

*Figure 18: Plot of AIC as the number of lags increase*

I aggregated the variables used in $VAR(33)$ as a $(7 \times 1)$ vector of time series variables:

$$Y_t = [\Delta yieldsp_t \ \ \Delta termpr_t \ \ \Delta forward1yr_t \ \ 1yr-ffr_t \ \ rec\_ind_t \ \ ted_t \ \ vix_t]'$$

$VAR(33)$ is a seemingly unrelated regression (SUR) model with $k = 7$ deterministic regressors and $l = 33$ lagged variables of the form:

$$\Delta yieldsp_t = \alpha_1 + \sum_{l=1}^{33} \beta_{1,l}^{\Delta yieldsp} \Delta yieldsp_{t-l} + \sum_{l=1}^{33} \beta_{2,l}^{\Delta yieldsp} \Delta termpr_{t-l} + \ldots + \sum_{l=1}^{33} \beta_{7,l}^{\Delta yieldsp} vix_{t-l} + \varepsilon_{\Delta yieldsp, t}$$

$$\Delta termpr_t = \alpha_2 + \sum_{l=1}^{33} \beta_{1,l}^{\Delta termpr} \Delta yieldsp_{t-l} + \sum_{l=1}^{33} \beta_{2,l}^{\Delta termpr} \Delta termpr_{t-l} + \ldots + \sum_{l=1}^{33} \beta_{7,l}^{\Delta termpr} vix_{t-l} + \varepsilon_{\Delta termpr, t}$$

$$\vdots$$

$$\Delta vix_t = \alpha_7 + \sum_{l=1}^{33} \beta_{1,l}^{vix} \Delta yieldsp_{t-l} + \sum_{l=1}^{33} \beta_{2,l}^{vix} \Delta termpr_{t-l} + \ldots + \sum_{l=1}^{33} \beta_{7,l}^{vix} vix_{t-l} + \varepsilon_{vix, t}, \text{ where:}$$

$\varepsilon_t = [\varepsilon_{\Delta yieldsp, t} \ \ \varepsilon_{\Delta termpr, t} \ \ \ldots \ \ \varepsilon_{vix, t}]'$ is a white noise (WN) process $(7 \times 1)$ vector of "innovations," which are forecast errors of a variable conditional on the lagged values of itself and those of other variables. $\varepsilon_t \sim WN(0, \Sigma)$. In the reduced-form VAR, the innovations don't have a structural interpretation, and errors of a variable are correlated with those of the other variables, possibly due to contemporaneous causal associations or common influence of other variables. For example, $corr(\varepsilon_{vix,\ t},\ \varepsilon_{\Delta yieldsp,\ t}) \neq 0$ as $cov(\varepsilon_{vix,\ t},\ \varepsilon_{\Delta yieldsp,\ t}) = \sigma_{vix,\ \Delta yieldsp}$. Due to this correlation, we can enhance the efficiency by using the SUR system estimator, than by individually estimating each equation separately by the OLS estimator.

The Durban−Watson test statistic value for all the variables in the VAR system is 2, implying that they are serially uncorrelated. Each variable is a linear function of lags 1 to 33, and all the other variables in the VAR system. Thus, all the 33 lags of each of the seven variables are the regression predictors for each of each of the seven variables. As a multivariate time series is the dependent variable, it can forecast all the variables in $Y_{t+h}$. So, given the data upto time $t$, we can forecast for time $t+1$ upto $t+h$ where $h > 1$ iteratively using OLS or a faster method called GLS. As I modeled $VAR(33)$ with difference $yieldsp$, after calibrating the rolling forecasts in the test set, I inverted $\Delta\ yieldsp$ to the leveled values − $invert\_yieldsp\_var$, which resulted in the RMSE of 0.4648.

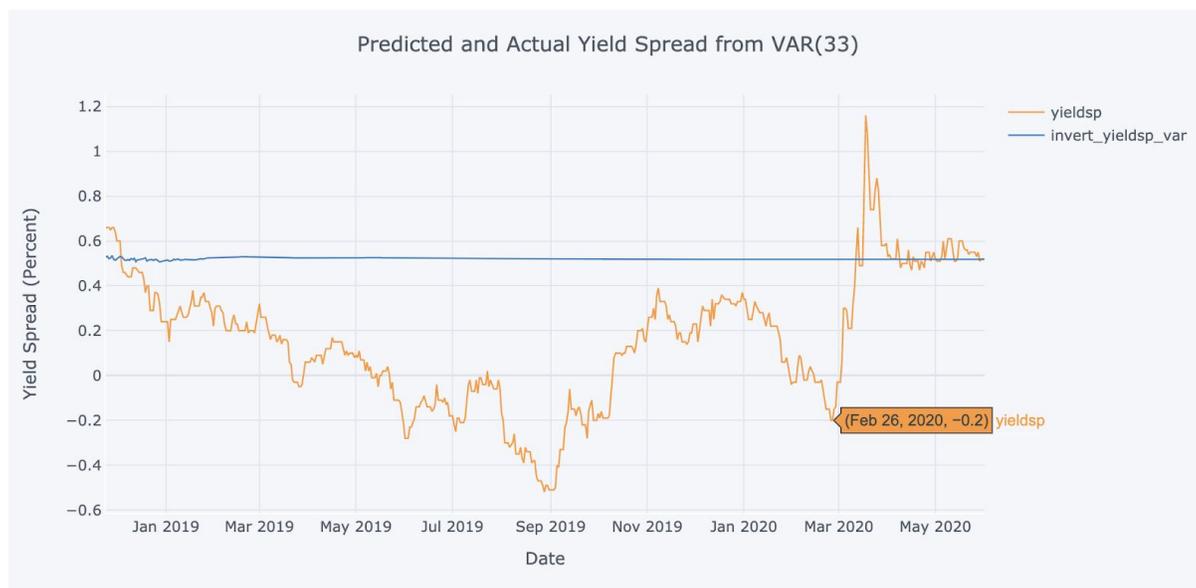

Figure 19: Forecasted and actual yieldsp from VAR(33)

The forecasts from figure 19 are slightly wavy prior to January 2019, prior to which they are relatively constant at 0.66, unable to capture the variations in movements from the peak to trough, possibly signifying high bias and unfitting the data. This contrasts with the other models that vary in conjunction with the ex-post values over time. A probable reason is that the VAR model is excessively parameterized as there are 7 variables each with 33 lags and constants, resulting in the model to estimate a total of 238 parameters. A dearth of information to determine the model's coefficients can culminate in diffusive predictive distributions and inaccurate predictions.

# Impulse Response Functions (IRF) and Forecast Error Variance Decomposition (FEVD)

I have examined the properties of VAR by doing structural analyses: impulse responses and forecast error variance decomposition. Individual estimates of coefficients only yield limited information about how a system reacts to a shock as all the variables in a VAR model are associated with each other. To obtain a border dynamic picture of the model's behavior, I have constructed graphs of impulse response functions. Impulse responses identify shocks to the VAR model. In the context of a VAR model, the IRFs trace out the time path of the effects of an exogenous shock $\varepsilon_t$ to one (or more) of the endogenous variables on some or all of the other variables in a VAR system given that no future innovations are present in the system. By back substitution, the standard MA representation of the simultaneous equations in $VAR(33)$ is as follows:

$$\Delta yieldsp_t = \varepsilon_{\Delta yieldsp, t} + \sum_{l=1}^{33} \beta_{1,l}^{\Delta yieldsp} \varepsilon_{\Delta yieldsp, t-l} + \sum_{l=1}^{33} \beta_{2,l}^{\Delta yieldsp} \varepsilon_{\Delta termpr, t-l} + \ldots + \sum_{l=1}^{33} \beta_{7,l}^{\Delta yieldsp} \varepsilon_{vix, t-l} + \ldots$$

$$\Delta termpr_t = \varepsilon_{\Delta termpr, t} + \sum_{l=1}^{33} \beta_{1,l}^{\Delta termpr} \varepsilon_{\Delta yieldsp, t-l} + \sum_{l=1}^{33} \beta_{2,l}^{\Delta termpr} \varepsilon_{\Delta termpr, t-l} + \ldots + \sum_{l=1}^{33} \beta_{7,l}^{\Delta termpr} \varepsilon_{vix, t-l} + \ldots$$

.
.
.

$$\Delta vix_t = \varepsilon_{vix, t} + \sum_{l=1}^{33} \beta_{1,l}^{vix} \varepsilon_{\Delta yieldsp, t-l} + \sum_{l=1}^{33} \beta_{2,l}^{vix} \varepsilon_{\Delta termpr, t-l} + \ldots + \sum_{l=1}^{33} \beta_{7,l}^{vix} \varepsilon_{vix, t-l} + \ldots$$

By Cholesky decomposition, we can normalize the MA representation for impulse response analysis, wherein the innovations of the transformed system are now in standard deviation units.

$$\Delta yieldsp_t = b_0^{\Delta yieldsp} \varepsilon'_{\Delta yieldsp, t} + \sum_{l=1}^{33} b_{1,l}^{\Delta yieldsp} \varepsilon'_{\Delta yieldsp, t-l} + \sum_{l=1}^{33} b_{2,l}^{\Delta yieldsp} \varepsilon'_{\Delta termpr, t-l} + \ldots + \sum_{l=1}^{33} b_{7,l}^{\Delta yieldsp} \varepsilon'_{vix, t-l} + \ldots$$

$$\Delta termpr_t = b_0^{\Delta termpr} \varepsilon'_{\Delta termpr, t} + \sum_{l=1}^{33} b_{1,l}^{\Delta termpr} \varepsilon'_{\Delta yieldsp, t-l} + \sum_{l=1}^{33} b_{2,l}^{\Delta termpr} \varepsilon'_{\Delta termpr, t-l} + \ldots + \sum_{l=1}^{33} b_{7,l}^{vix} \varepsilon'_{vix, t-l} + \ldots$$

.
.
.

$$\Delta vix_t = b_0^{vix} \varepsilon'_{vix, t} + \sum_{l=1}^{33} b_{1,l}^{vix} \varepsilon'_{\Delta yieldsp, t-l} + \sum_{l=1}^{33} b_{2,l}^{vix} \varepsilon'_{\Delta termpr, t-l} + \ldots + \sum_{l=1}^{33} b_{7,l}^{vix} \varepsilon'_{vix, t-l} + \ldots, \text{ where:}$$

$\varepsilon_{\Delta yieldsp, t} \sim WN(0, 1), \ \varepsilon_{\Delta termpr, t} \sim WN(0, 1), \ \ldots, \ \varepsilon_{vix, t} \sim WN(0, 1)$

The normalization yields zero covariance between the disturbances of the transformed system. Consequently, we can shock a variable at a time, isolating the effects of other variables. I have computed the response of $\Delta yieldsp_t$ to a unit normalized innovation to all the variables in the $Y_t$ matrix. At time horizon $h$, the impulse responses of variables due to an exogenous shock to $\Delta yieldsp_t$ is the derivative with respect to the shocks. Using Cholesky decomposition, we decompose the variance-covariance matrix $\Sigma$ into a lower triangular matrix $P$ and its transpose $P'$ such that $\Sigma = P\,P'$ where the diagonal elements of $P$ are positive.

$$Imp_{\Delta yieldsp}(h) = \{\frac{d\Delta yieldsp_{t+h}}{d\varepsilon'_t}\}^{10}_{h=0} = B^h P, \text{ where } B \text{ is the coefficient matrix in the VAR model.}$$

Figure 20 traces the impact of shocks of macro variables (present in the VAR system) onto the difference *yieldsp*. A shock of $\Delta\,termpr$ does not affect $\Delta\,yieldsp$ till the second period and raises the $\Delta\,yieldsp$ to estimated 0.45 in period four. Thereafter, it gradually slumps to the negative territory at about $-0.25$ before converging to 0 in the tenth period. An impulse response from $\Delta\,forward1yr$ has somewhat opposite effect due to the near reverse in the trajectory of $\Delta\,yieldsp$ over time. As $\Delta\,yieldsp$ is stationary, the impulse response of shocks on $\Delta\,yieldsp$ decays to 0 by the tenth period, signifying that one-time innovations don't have long-term ramifications on the paths of $\Delta\,yieldsp$.

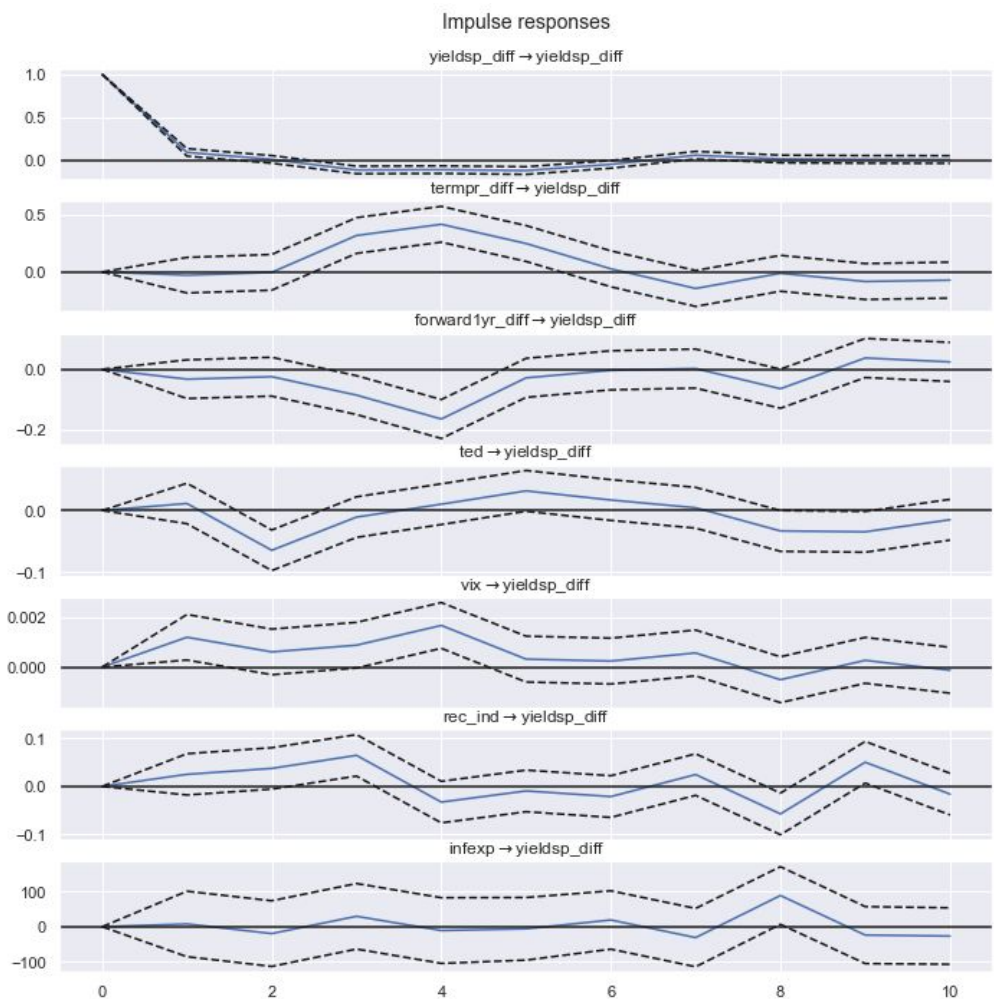

*Figure 20: Impulse responses functions depicting shocks from the exogenous variables to yieldsp*

Another way to characterize the dynamics associated with the VAR is by computing the FEVD from the VAR model. We can find how much a shock to one variable such as *ted, vix,* etc impacts the variance of the forecast error of a different one, such as $\Delta yieldsp$. So, in the short run, i.e. in the same period, $\Delta yieldsp$ explains 59.78 percent of the variance in forecast error of $\Delta termpr$ − this strong influence on $\Delta termpr$ indicates that it is strongly endogenous. *ted*, *vix*, $\Delta forward1yr$ and *rec_ind* are strongly exogenous as they only weakly influence in predicting $\Delta termpr_t$. $\Delta termpr_t$ itself explains 40.21 percent of variance in its error. However, in the long run, the influence of $\Delta termpr$ in explaining the variance in its forecast error diminished marginally from 40.21 to 38.58 percent, whereas $\Delta yieldsp$ has an incremental influence, as its effect rises from 59.78 to 60.85 percent in the short in the earlier periods in the short run.

**Multilayer Perceptron (MLP)**

Shifting gears from the classical econometrics forecasting tools, now I have explored the forecasting mechanisms from machine learning − *MLP* and *LSTM*. One of the simplest architectures of neural networks is the multilayer perceptron (MLP). On a high level, MLP consists of:

Input layer: vector of independent variables or features

Hidden layer − $h_t$ : each hidden layer consists of *n* neurons

Output layer: outputs the network and displays the prediction.

First, we transform the input of each hidden layer linearly and apply the rectified linear (ReLU) activation function − $f(x) = max(0, x)$, to transform the input non-linearly. This non-linear transformation enables MLP to identify complex non-linear relationships between *yieldsp* and the independent variables with missing values, and is robust to noise. However, the caveat is that we can only provide a fixed number of inputs to generate a fixed number of outputs by enumerating the temporal dependence in the model's design. The firm mapping function between the inputs and the outputs in the feedforward neural network is problematic when we feed in a sequence of inputs in the model.

Prior to fitting the model in the train set, I transformed the time series into a supervised learning problem where the observations in the previous time step : $yieldsp_{t-1}$ become inputs to forecast $yieldsp_t$. Moreover, I rescaled the data such that $yieldsp_t \in (-1, 1) \lor t.$ Increasing the number of lags as explanatory variables or inputs, I have simultaneously expanded the network's capacity by varying the number of neurons in a single hidden layer $(1, 3, 5)$, albeit at a possible risk of overfitting the training data. Automatically, incremental lags scale the input neurons. For instance, 3 input neurons stem from 3 lags. Thereby, I have also scaled the number of neurons in the hidden layer by inputting the equivalent number of lags in the hidden layer. With 3 lag observations − $yieldsp_{t-1}$, $yieldsp_{t-2}$, $yieldsp_{t-3}$, as inputs, I have used the same number of neurons, respectively. Changing the number of inputs alters the total number of training patterns when we convert the time series data into a supervised learning problem. Furthermore, I have conducted experiments with a batch size to 2. Consisting of 20 training epochs in each experimental run, every experimental scenario is run 8 times and I have recorded the RMSE after each run ends. From highest to lowest, the test RMSE ranges from 0.056 to 0.054, with RMSE generally reducing after each experimental run. The descriptive statistics in table 5 shows how the test set RMSE changes with different numbers of neurons in the hidden layer.

|  | 1 neuron | 3 neurons | 5 neurons |
| --- | --- | --- | --- |
| Count | 8 | 8 | 8 |
| Mean | 0.055470 | 0.055397 | 0.054927 |
| Standard Deviation | 0.000165 | 0.000388 | 0.000445 |
| Minimum | 0.055119 | 0.054996 | 0.054463 |
| $25^{th}$ Percentile | 0.055447 | 0.054996 | 0.054463 |
| $50^{th}$ Percentile | 0.055452 | 0.055373 | 0.054941 |
| $75^{th}$ Percentile | 0.055616 | 0.055427 | 0.055319 |
| Maximum | 0.055617 | 0.056244 | 0.055432 |

Table 5. Descriptive statistics of the test RMSE with different neurons in the hidden layer in MLP

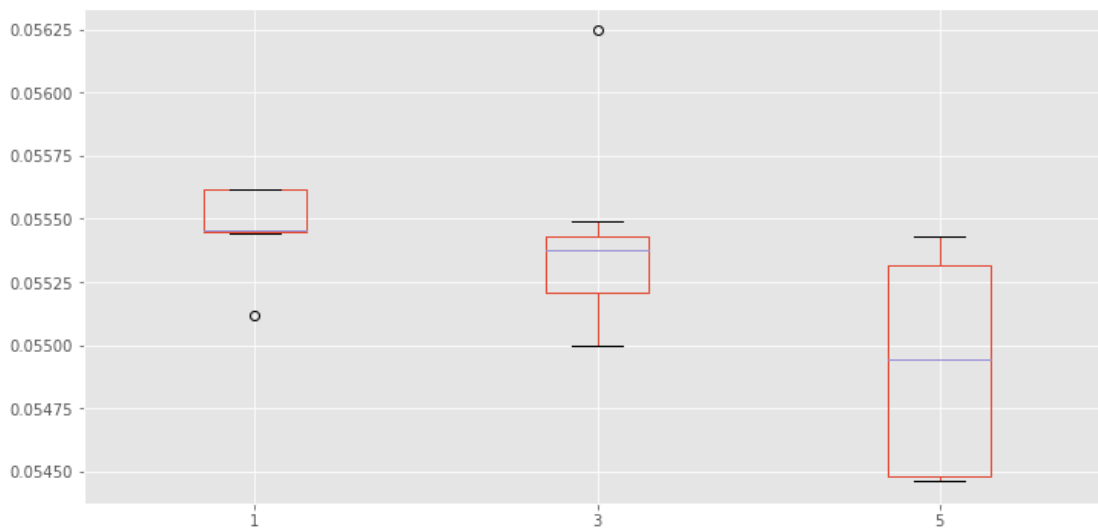

Figure 21: Box plot that shows the lowest RMSE occurs with 5 neurons in the hidden layer in MLP.

I have run the configuration for a total of 10 times. To check if the model's configuration overfits or undefits the training data, I have evaluated the RMSE on both the train and test data after each of the 20 epochs and plotted them in a line plot in figure 22. The results indicate that the model runs well without overfitting as the line plot of the test RMSE (yellow color line) is lower than the train RMSE (blue color line).

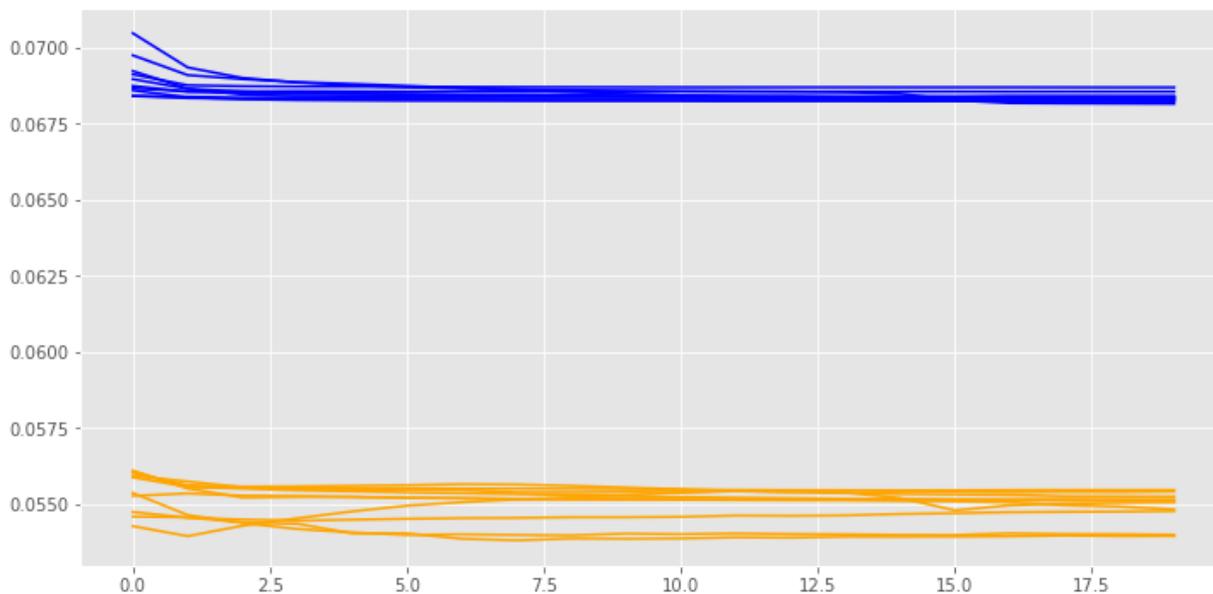

Figure 22: Line plot that highlights lower test RMSE relative to the train RMSE through each of the 20 epochs.

Unlike regression predictive techniques like ARIMA that don't consider the complexity arising from sequence dependence among the input variables, recurrent neural networks (RNN) are powerful deep learning methods to manage sequence dependence. RNNs are called recurrent as they perform the same task for every element of a sequence, and the output i.e. the ex-ante *yieldsp* depends on the new input i.e. $yieldsp_{t+1}$ and all the historical values prior to $yieldsp_{t+1}$ fed in the past. While a vanilla neural network is usually very constrained as it accepts a fixed-sized vector as inputs and produces a fixed-size vector as output, a type of RNN is the Long Short-Term Memory (LSTM) can capture long−term dependencies than vanilla RNNs. Its convoluted architecture can be successfully trained using Backpropagation Through Time, wherein overcomes the problem of vanishing gradients.

**Long Short Term Memory (LSTM)**

In contrast with the neurons in vanilla artificial neural networks (ANN), the LSTM networks have memory blocks (or cell states) linked through layers. As shown in the diagram, the cell state $C_{t-1}$ is a horizontal line that runs on top of the diagram. Like a conveyor belt, it flows down the entire chain, linearly interacting only a few times. The "gates" are structures that optimally regulate the flow (addition or subtraction) of information in the LSTM's cell states. The components of a block enable LSTM to work smarter than ANNs as it acts as a repository of memory for the most recent sequence. Furthermore, a cell state consists of gates that regulate the block's state and output. Operating on a given sequence of input, each gate within the block uses the sigmoid activation function to control if they are triggered or not. If triggered, the state changes and information flows through the cell state conditionally.

A unit, like a mini-state machine, comprises of three types of gates as depicted in figure 23:

**Forget Gate:** It decides the information to throw away from the cell state. For each number in the cell state $C_{t-1}$, the output from the forget layer is $f_t = \sigma [w_f (h_{t-1}, x_t) + b_f] = \sigma (w_{fx} x_t + h_t w_{fh} + b_f)$

Here, we apply the sigmoid function of the form: $\sigma(x) = \frac{1}{1+e^{-x}} \in [0, 1]$, and the output from the sigmoid layer describes how much information would be let through the gate. $h_t$ is the hidden state at time step $t$, also known as the "memory" of the network as it captures information about the sequence of events that occurred in all the previous time steps. It is the output of the current cell. Together the weight matrices — $w_h$ and $w_x$ define how to calculate the network's new memory given the previous input and memory. The assigned weights are updated or learned while training the model. If $\sigma(x) = 1$, then everything passes through the data; however, if $\sigma(x) = 0$, then nothing passes.

**Input Gate:** It is a sigmoid layer that decides the values from the input to update the memory state.

$i_t = \sigma [w_i (h_{t-1}, x_t) + b_i] = \sigma (w_{ix} x_t + h_t w_{ih} + b_i)$

Then, a layer with the tangent hyperbolic activation function of the form: $tanh(x) = \frac{e^x - e^{-x}}{e^x + e^{-x}} \in [-1, 1]$, creates a vector of new candidate values $C'_t$ that could be added to the state.

$C'_t = tanh [w_c( h_{t-1}, x_t) + b_c] = tanh (w_{cx} x_t + w_{ch} h_{t-1} + b_c)$. The *tanh* function distributes gradients such that the cell state $C_t$ information flows longer without vanishing or exploding.

Subsequently, we update the old cell state $C_{t-1}$ to a new cell state $C_t$. Multiplying the old state $C_{t-1}$ by $f_t$, forgets the information the model forgot earlier. Afterwards, we add new candidate values scaled by the magnitude the model decided to update the value in each state — $iC'_t$. Therefore, the complete equation to update the cell state is
: $C_t = f_t C_{t-1} + i_t C'_t$

**Output Gate:** As before, a sigmoid layer that decides the output based on the cell state's memory and input.

$\widehat{y}_t = \sigma [w_y (h_{t-1}, x_t) + b_y] = \sigma (w_{yx} x_t + h_t w_{yh} + b_y)$

Finally, the cell state state passed through *tanh* function to squeeze the values between $-1$ and 1 as follows:

$h_t = \widehat{y}_t \, tanh (C_t)$

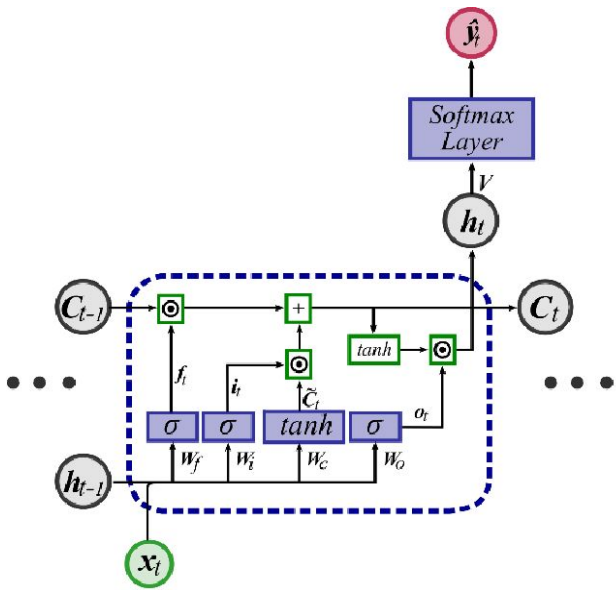

*Figure 23: A block diagram that describes the cell unit of LSTM recurrent neural networks. [5]*

I have constructed a multi-layered LSTM network to forecast the test set values of *yieldsp*, given the history of observations in the training set. As LSTMs are sensitive to the scale of the input data, particularly when we invoke the sigmoid or the hyperbolic tangent activation functions, I have normalized or rescaled the data to the range $(0, 1)$. This model runs under "stateless case", wherein the model updates the parameters for each batch *i,* but then initiates the cell and hidden states to 0 for the subsequent batches − $i + 1$, $i + 2, ...$

In developing the LSTM model's architecture, I applied a regularization technique called "dropout" wherein a proportion of recurrent connections and inputs to the LSTM units are probabilistically excluded from the process of updating weights and activation when we train a network. In this case, I've dropped 20 percent of the data and trained the network for 1000 epochs. This diminishes overfitting and enhances the model's performance. Overfit models that occur in convoluted models tend to describe random noise in the data, instead of explaining the true relationship between variables, raising the variance but reducing the bias. Setting a batch size of 100 propagates 100 observations (in the train set) chronologically through the network each time and fits the model with the observations in the batch size. Additionally, we can enhance the model's performance by tuning the hyperparameters such as the type of optimizer. This is crucial as ideally the optimizer should reach global minima where the cost function is the lowest. Thus, I have compiled the model using the iterative method called the RMSprop optimizer. Another regularization tool that I have applied using a callback function is "early stopping." This updates the learner to ensure that the training data is fit better at each iteration and avoids overfitting. Usually, a model may perform well upto a certain point i.e. the loss is low; whereas it raises past that point. Therefore, early stopping guides the number of iterations to run before overfitting the model. Specifically, the callback function monitored the loss in the validation (test) set at each epoch, and training is interrupted if the test set loss does not improve after 10 epochs. If the test loss drops below training loss, then the model may be overfitting the training data. Whilst I allowed the network to train for 1000 epochs, the model converged to the optimal value in 207 epochs. The LSTM model resulted in the train and test RMSE as 0.0982, and 0.0630, respectively, implying that the model is not overfit. Measuring and plotting the test and train RMSE in figure 24 shows that the loss in the validation set — $val\_loss,$ is lower than than the train set across the 287 epochs.

---

[5] The diagram is taken from:
https://www.researchgate.net/figure/Block-diagram-of-the-LSTM-recurrent-neural-network-cell-unit-Blue-boxes-means-sigmoid_fig2_328761192

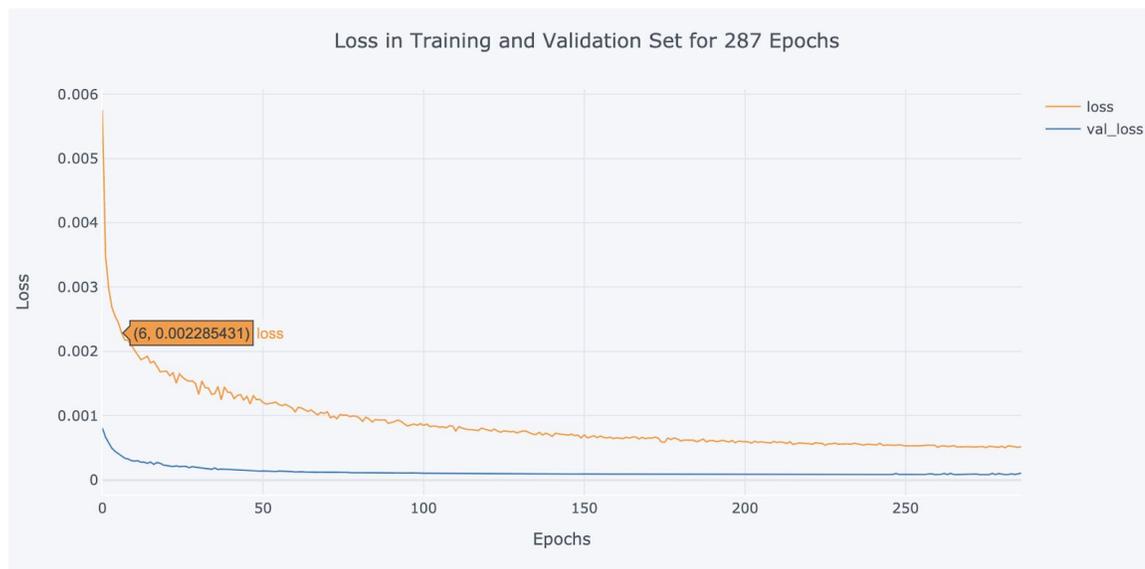

*Figure 24: Loss in train and test set across 287 epochs*

Normally, the state within the network is reset after each training batch when fitting the model as done in the previous "stateless" LSTM model. However, by making the layer "stateful", we can gain finer control over the internal state of the network by building stacked LSTMs. Here, the output from the hidden and cell state of batch $i$ becomes the input for batch $i + 1$ as memory in LSTMs enable the network to automatically recall across longer temporal dependence. Stacking is the process of building the layers of LSTM such that the output of one layer : $h_l : l = t,\ t + 1,\ t + 2, ...$ becomes the input of another layer, making the model deeper. This means that it can build state over the entire training sequence and even maintain that state if needed to make predictions. This model yielded a train and test set RMSE of 0.07999, and 0.06547, respectively.

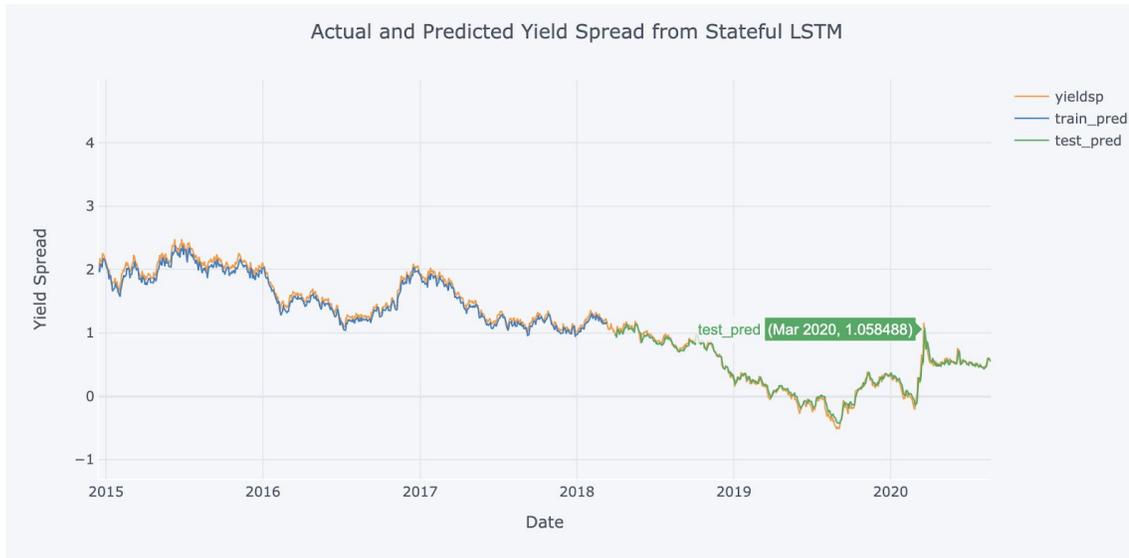

*Figure 25: Ex-post and ex-ante yieldsp from the Stateful LSTM for train and test set.*

The unsophisticated linear models such as ARIMA cannot learn a long-term structure embedded in the data. For instance, an *AR(p)* cannot learn dependencies greater than *p,* warranting the series to be locally stationary. Without stationarizing the variables, forecasts become behave bizarrely for large values of the time horizon *h,* and the variance of the forecast errors may explode if $h \to \infty$. Unlike these linear models, LSTMs can learn non-linearities very well as a larger dataset facilitates them to learn long-term dependencies. Hence, in constructing LSTMs using multivariate data, I used all the variables as given, without stationarizing the variables with unit roots. First, splitting the dataset into train and test sets, then I further divided the train and test sets into input and output variables. Thereafter, I defined the LSTM wherein the first hidden layer comprises 25 neurons and the output layer has 1 neuron. Optimizing the model using the RMSProp method, the model computes the loss function via the mean absolute error. Fitting the model for 500 training epochs with a batch size of 50, the LSTM's internal state is reset at the end of each batch size. Thus, the internal state is a function of the number of observations. Figure 26 tracks both the train and test loss for upto 500 epochs. Till the first 25 epochs, the loss in the test set rises as is greater than the train set loss, indicative of possible overfitting. However, it gradually receded and falls below the train loss after the $417^{th}$ epoch, nullifying the effects and concerns of an overfit model.

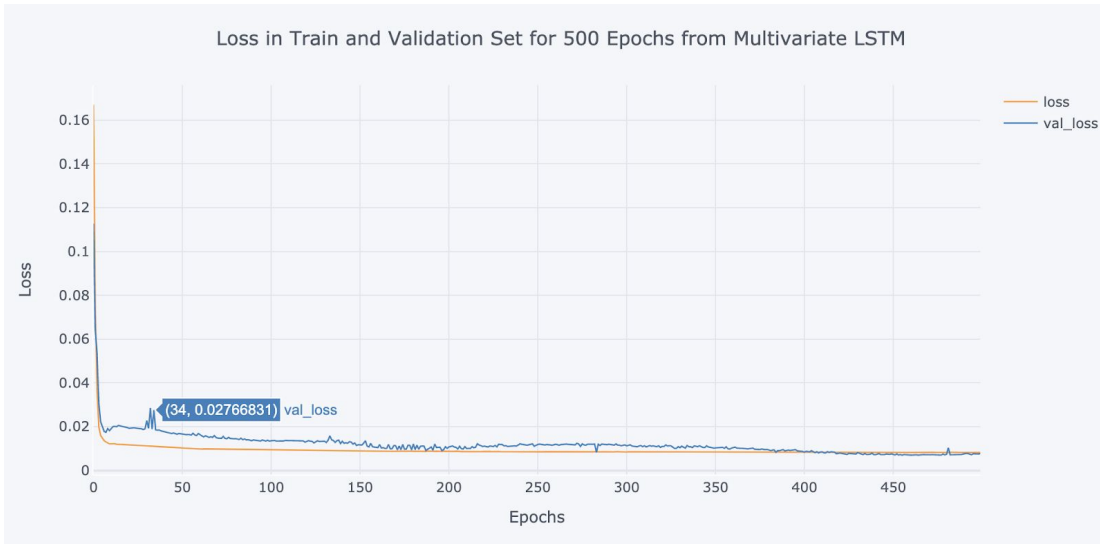

*Figure 26: Loss in train and test set across 500 epochs from multivariate LSTM*

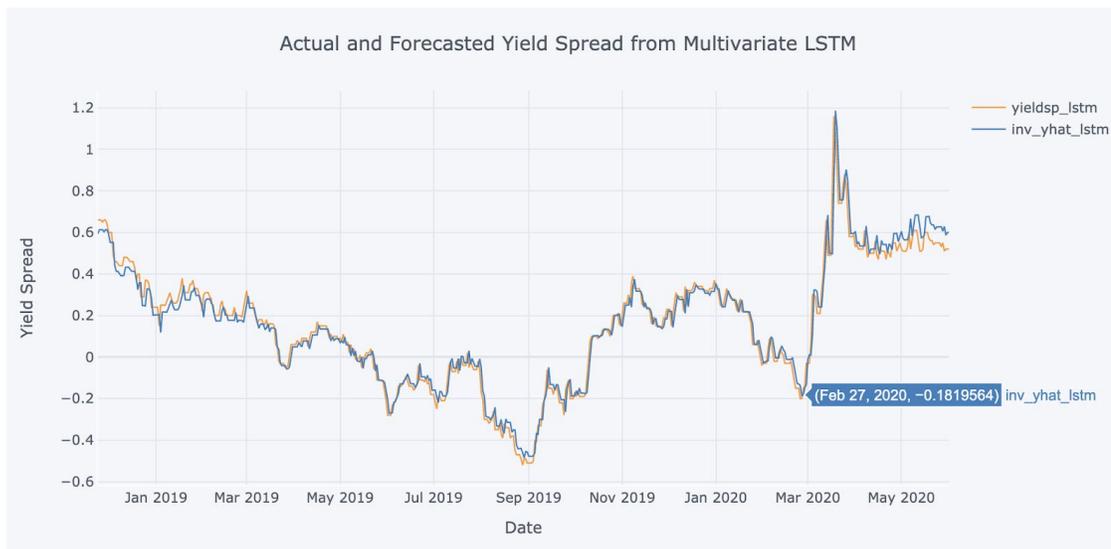

*Figure 27: Ex-post (yieldsp_lstm) and ex-ante (inc_yhat_lstm) yieldsp for the test set from the multivariate LSTM model.*

Finally, I combined the forecast with the test data and inverted the scaling on both the training and test sets. After the ex-ante and ex-post values are on the original scale, I calculated the test RMSE for the model, which outputs the error in the same units as the variable *yieldsp* itself. The test RMSE of 0.05185 indicates an improvement from the univariate predictions.

## 4. Conclusion and Discussion

The emphasis of this paper is to contrast the classical and emerging forecasting techniques and using the metric of root mean squared error to evaluate which model produces the best results. Table 6 compares the test RMSE obtained from each model, from which I concluded that predictions from $VAR(33)$ and multivariate LSTM lie on the opposite end of the spectrum, with the former performing relatively poorly than the latter. The unchanged ex-ante *yieldsp* from the highly parameterized reduced form $VAR(33)$ starkly contrasts with those from all other models which are able to capture the changing patterns. A possible way to improve the model is by developing a structural VAR, where the innovations have structural interpretations as deeper structural shocks drive the innovations. Alternatively, we could develop a vector error correction method that represents a cointegrating VAR. Besides being more efficient than the VAR, it explains the long and short term relationship between the variables and we can evaluate how to correct deviations from the long run.

| Model | Test RMSE |
| --- | --- |
| $ARIMA(1,0,3)$ | 0.05185 |
| $Difference\ ARIMA(3,1,3)$ | 0.05615 |
| $SARIMAX(2,1,2)$ | 0.3458 |
| $VAR(33)$ | 0.4648 |
| $MLP$ | 0.0544 |
| $Stateful\ LSTM$ | 0.06547 |
| $Multivariate\ LSTM$ | 0.05185 |

*Table 6. Summary of the test RMSE from all the models*

Similarly, one could construct a Bayesian VAR, build on the fundamentals of Bayes Theorem. Whilst the reduced form parameters are passed in the likelihood function, the orthogonal matrix $\Omega$ does not enter the likelihood function. Therefore, we cannot identify $\Omega$ based on the given sample. Consequently, the conditional posterior will be identical to the conditional prior as the conditional distribution of the parameters in reduced-form will not be updated. Comparatively, identifications don't hinder the conceptual workings of the Bayesian analysis to a great extent. In Bayesian inference, the sample contained in the likelihood function updates the prior to form a posterior distribution. So, we can assign probabilities to the model specifications and update those values after observing the data. It is crucial to know the predictive distributions of ex-ante values such as inflation rate, yield curve and other macroeconomic variables relevant for policy-making. Equally essential is accounting for the uncertainty about the actual shocks and the estimated parameters. As Bayesian methods treat these parameters and shocks symmetrically as random variables, it is straightforward to simultaneously consider these two sources of uncertainty. Finally, we can explore a related methodology called Gaussian Process, as it hinges on multivariate Gaussian distribution.